\documentstyle[12pt,epsfig,color]{article}
\newcommand{\be}{\begin{equation}}
\newcommand{\ee}{\end{equation}}
\newcommand{\beq}{\begin{equation}}
\newcommand{\eeq}{\end{equation}}
\newcommand{\ba}{\begin{eqnarray}}
\newcommand{\ea}{\end{eqnarray}}
\newcommand{\bea}{\begin{eqnarray}}
\newcommand{\eea}{\end{eqnarray}}
\newcommand{\nn}{\nonumber}
\begin{document}
\baselineskip=15.5pt \pagestyle{plain} \setcounter{page}{1}
%
\begin{titlepage}

\vskip 0.8cm

\begin{center}

{\Large \bf  Towards 1/N corrections to deep inelastic scattering
from the gauge/gravity duality}

\vskip 1.cm

{\large {{\bf David Jorrin}{\footnote{\tt
jorrin@fisica.unlp.edu.ar}}, {\bf Nicolas Kovensky}{\footnote{\tt
nico.koven@fisica.unlp.edu.ar}}, {\bf and Martin
Schvellinger}{\footnote{\tt martin@fisica.unlp.edu.ar}}}}

\vskip 1.cm

{\it IFLP-CCT-La Plata, CONICET and Departamento  de F\'{\i}sica,
Universidad Nacional de La Plata.  Calle 49 y 115, C.C. 67, (1900)
La Plata,  Buenos Aires, Argentina.} \\

\vspace{1.cm}

{\bf Abstract}

\vspace{1.cm}

\end{center}

$1/N^2$ corrections to deep inelastic scattering (DIS) of charged
leptons from glueballs at strong coupling are investigated in the
framework of the gauge/gravity duality. The structure functions
$F_1$ and $F_2$ (and also $F_L$) are studied at subleading order in
the $1/N^2$ expansion, in terms of $q^2$ and the Bjorken parameter
$x$. The relevant type IIB supergravity one-loop diagrams (which
correspond to DIS with two-hadron final states) are studied in
detail, while $n$-loop diagrams (corresponding to DIS with
$(n+1)$-hadron final states) are briefly discussed. The $1/N^{2n}$
and $\Lambda^2/q^2$ dependence of the structure functions is
analyzed. Within this context two very different limits are
considered: one is the large $N$ limit and the other one is when the
virtual photon momentum transfer $q$ is much larger than the
infrared confining scale $\Lambda$. These limits do not commute.

\noindent

\end{titlepage}

\newpage

{\small \tableofcontents}

\newpage

\section{Introduction}

The idea of the present work is to investigate $1/N^2$ corrections
to DIS of charged leptons off glueballs at strong coupling by using
the gauge/gravity duality\footnote{$N$ is the rank of the gauge
group of the gauge theory.}. This corresponds to a DIS process where
there are two-hadron final states. By using the optical theorem this
is related to a forward Compton scattering (FCS) process with
two-particle intermediate states, {\it i.e.} one-loop FCS Feynman
diagrams. Moreover, we also consider $1/N^{2n}$ corrections to DIS
(where $n$ is an integer), which corresponds to $(n+1)$-hadron final
states, while in terms of FCS it is related to $(n+1)$-particle
intermediate states, {\it i.e.} $n$-loop FCS Feynman diagrams.

In terms of the gauge/string duality Polchinski and Strassler
studied scattering processes in the large $N$ limit both for hard
scattering \cite{Polchinski:2001tt} and for DIS
\cite{Polchinski:2002jw}. Further work related to DIS from the
gauge/string duality includes
\cite{Brower:2006ea,Cornalba:2006xk,Cornalba:2006xm,Cornalba:2007zb,Brower:2007qh,Brower:2007xg,Cornalba:2008qf,
Cornalba:2008sp,BallonBayona:2009uy,Cornalba:2009ax,Cornalba:2010vk,Koile:2011aa,Koile:2013hba,Gao:2014nwa,
Brower:2014wha,Koile:2014vca,Koile:2015qsa,Capossoli:2015sfa,Capossoli:2015ywa}.
For DIS in \cite{Polchinski:2002jw} the authors considered the
structure functions of glueballs in the case when there is a
single-hadron final state. In addition, they suggested that for
two-hadron final states DIS can also be studied within the
supergravity description. Thus, we will investigate type IIB
supergravity loop corrections, in particular describing in detail
one-loop corrections.

\begin{figure}[h]
\begin{center}
\includegraphics[scale=0.35]{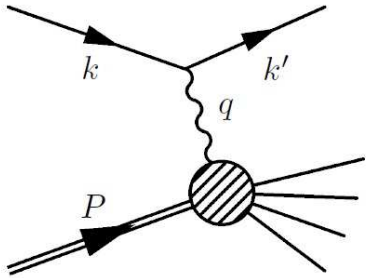}
\label{DIS} \caption{\small Schematic picture of DIS of a charged
lepton with four-momentum $k$ off a hadron with four-momentum $P$,
through the exchange of a virtual photon with four-momentum $q$.}
\end{center}
\end{figure}

DIS of a charged lepton off a hadron is schematically shown in
figure \ref{DIS}.
The process involves a charged lepton with four-momentum $k$, which
emits a virtual photon with four-momentum $q$. This probes the
internal structure of a target hadron with initial four-momentum
$P$. The scattering cross section of DIS is proportional to the
contraction of a leptonic tensor, $l_{\mu\nu}$, described by using
perturbative QED, and a hadronic tensor, $W_{\mu\nu}$, which is
difficult to calculate since it involves soft QCD processes. At weak
coupling, the parton model describes this process: the virtual photon
interacts directly with one of the partons inside the hadron. At
strong coupling, on the other hand, the parton model is not a
suitable description and therefore a different strategy must be considered.
We will use an approach based on the gauge/string duality and the methods
developed in \cite{Polchinski:2002jw}.

In general terms, from the theoretical point of view, there is a
standard way to proceed in order to study the internal structure of
hadrons. In fact, by using the optical theorem the DIS cross section
is related to the matrix element of a product of two electromagnetic
currents $J^\mu(x) \, J^\nu(0)$ inside the hadron, which corresponds
to the FCS process\footnote{$J^\mu(x) \, J^\nu(0)$ correlation
functions have also been calculated at strong coupling for the
${\cal {N}}=4$ SYM theory plasma, both in the DIS regime
\cite{Hatta:2007he,Hatta:2007cs} and in the hydrodynamical one
\cite{CaronHuot:2006te}. Also, the corresponding leading string
theory corrections (${\cal {O}}(\alpha'^3)$, with $\alpha'=l_s^2$),
which allow one to investigate the strong coupling expansion in
powers of $1/\sqrt{\lambda}$ (where $\lambda$ is the 't Hooft
coupling) in the gauge theory, have been calculated in both regimes
in \cite{Hassanain:2009xw} and
\cite{Hassanain:2010fv,Hassanain:2011fn,Hassanain:2011ce,Hassanain:2012uj},
respectively.}. The product of these two currents can be written in
terms of the operator product expansion (OPE), for an unphysical
kinematical region ({\it i.e.} for $x \gg 1$). Then, by using
dispersion relations it is possible to connect the above unphysical
result with the physical DIS cross section. The matrix element of
two electromagnetic currents inside the hadron is given by the
tensor $T^{\mu\nu}$, which is defined as
\begin{equation}
T^{\mu\nu}=i \int d^4x \, e^{iq\cdot x} \, \langle P, h'|{\hat T}(
J^{\mu}(x) \, J^{\nu}(0))|P,h \rangle \, ,
\end{equation}
where $h$ and $h'$ label the polarizations of the initial and final
hadronic states. $\hat T$ indicates the time ordered product of the
two currents. This tensor depends on $q^2$ and the Bjorken parameter
defined as
\begin{equation}
x=\frac{-q^2}{2 P \cdot q} \, ,
\end{equation}
being $0 \leq x \leq 1$ its physical kinematical range, where $x=1$
corresponds to elastic scattering. Beyond the physical kinematical
region, {\it i.e.} for $x > 1$, it is possible to carry out the OPE
of the tensor $T^{\mu\nu}$. This tensor is related by the optical
theorem to the hadronic tensor
\begin{equation}
W_{\mu\nu}(P,q)=i \int d^4x \, e^{iq\cdot x}\langle P,
h'|[J_{\mu}(x),J_{\nu}(0)]|P,h \rangle \, .
\end{equation}
Since we will focus on scalar glueballs, the hadronic tensor is
given by
\begin{equation}
W_{\mu\nu}= F_1(x,q^2) \left(\eta_{\mu\nu} - \frac{q_\mu
q_\nu}{q^2}\right) + \frac{2 x}{q^2} \, F_2(x,q^2) \left(P_\mu +
\frac{q_\mu}{2x} \right)\left(P_\nu + \frac{q_\nu}{2x} \right) \, ,
\label{tensorW}
\end{equation}
where $F_1(x,q^2)$ and $F_2(x,q^2)$ are the structure functions.
Recall that in the context of the parton model they are associated
with the distribution functions of the partons inside the hadron,
leading to the probability of finding a parton which carries a
fraction $x$ of the target hadron momentum, {\it i.e.} $x P$.

The optical theorem implies that $2 \pi$ times the imaginary part of
the structure functions associated with FCS gives exactly the DIS
structure functions. It allows one to calculate DIS structure
functions at strong coupling from the holographic dual description
given in \cite{Polchinski:2002jw}. In that paper a prescription for
the calculation of $W^{\mu\nu}$ for $1 \ll \lambda \ll N $, in the
planar limit of the gauge theory, has been developed. The idea is to
calculate the amplitude of a supergravity scattering process in the
bulk that turns out to be dual to the FCS in the boundary Yang-Mills
theory. According to that prescription, the insertion of a current
operator on the boundary induces a $U(1)$ metric
perturbation\footnote{Recall that the isometry group of $S^5$ is
$SO(6)$ which is related to the $SU(4)_R$ R-symmetry group of the
${\cal {N}}=4$ SYM theory. The idea is that the mentioned $U(1)$
group is a subgroup of $SO(6)$. Thus, the metric perturbation is
parameterized by an Abelian gauge field $A^\mu$ times a Killing
vector on $S^5$.}, that interacts with the dual type IIB
supergravity field of the glueball, {\it i.e.} the dilaton $\phi$.
The holographic picture is schematically depicted in figure 2.
\begin{figure}[h]
\begin{center}
\includegraphics[scale=0.5]{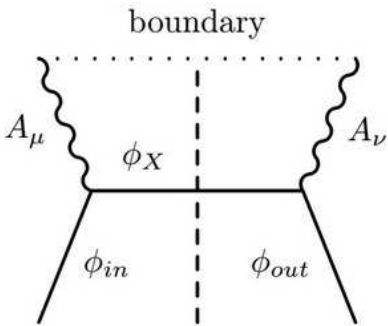}
\label{Schannel} \caption{\small Witten diagram representing the
$s$-channel contribution to the holographic dual description of FCS
in the $N\rightarrow \infty$ limit. The vertical dashed line
schematically represents the optical theorem procedure used to
extract the relevant imaginary part of $T^{\mu\nu}$. There is a
single-hadron intermediate state, which means that only
single-hadron final states are considered in DIS. The horizontal
dotted line represents the boundary of AdS$_5$. In addition,
$\phi_{in}$, $\phi_X$ and $\phi_{out}$ are the dual supergravity
fields of the initial, intermediate and final hadronic states,
respectively, while $A_\mu$ couples to $J^\mu$ of the quantum field
theory at the AdS boundary.}
\end{center}
\end{figure}
The sum over all possible on-shell intermediate states leads to a
formula for the imaginary part of the amplitude, and allows one to
obtain $F_1$ and $F_2$ and, from it, the longitudinal structure
function $F_L$. In this case in the FCS there is only one
intermediate state, which means that in the DIS that we consider
there is only one outgoing single-hadron final state. Note that
supergravity provides and accurate description of the holographic
dual process of DIS only if $\lambda^{-1/2} \ll x <1$. This is
because in this regime the Mandelstam variable $s$ (associated with
the center-of-mass energy) is not large enough in order to produce
excited string states. When $x$ becomes smaller than
$\lambda^{-1/2}$ it is necessary to consider the full string
theoretical description. On the field theory side, for $1 \ll
\lambda \ll N $ double-trace operators dominate the OPE. In fact two
very different kinds of limits can be considered, namely, the large
$N$ limit and the $q^2 \gg \Lambda^2$ limit, being $\Lambda$ the IR
confinement scale of the dual SYM theory.

It is very interesting to consider the case when DIS involves
two-hadron states as the final states. In this case the optical
theorem dictates that the holographic dual description of FCS is
given in terms of one-loop Witten diagrams, which in practical terms
are one-loop Feynman diagrams in type IIB supergravity. In fact, in
\cite{Polchinski:2002jw} it was suggested that this process can be
calculated by using supergravity. It gives the first correction to
DIS in the $1/N^2$ expansion. Also from each supergravity Feynman
diagram it is possible to extract the dependence in powers of
$\Lambda^2/q^2$. It is very interesting the fact that by taking
first the $N \rightarrow \infty$ limit, followed by the $q^2 \gg
\Lambda^2$ limit, it gives a totally different result compared with
the one obtained by taking these limits the other way around. This
effect has already been noted in a recent paper by Gao and Mou
\cite{Gao:2014nwa}, where this question has been addressed only in
part, by using an effective interaction Lagrangian in five
dimensions. On the other hand, in our present work instead we start
from the type IIB supergravity action in ten dimensions, including
all the relevant fields, thus carrying out a {\it first principles}
top-down calculation. These bulk fields correspond to specific
operators of the boundary SYM theory, which in this case is an IR
deformation of $SU(N)$ ${\cal {N}}=4$ SYM. In addition, we carry out
the explicit calculation of all relevant $t$-channel diagrams in
type IIB supergravity at leading order in $\Lambda^2/q^2$, taking
into account all possible Kaluza-Klein states within the full AdS$_5
\times S^5$ solutions of the bulk fields. In comparison with our
calculations, in \cite{Gao:2014nwa} only a few Kaluza-Klein states
have been considered, rendering their result incomplete in that
sense. Furthermore, we obtain the explicit functional dependence on
the Bjorken parameter at leading order in $\Lambda^2/q^2$. We find
that this dependence is consistent with the expectations of
\cite{Polchinski:2002jw} concerning the $1/N^2$ corrections.

Another new finding from our investigation is that it is interesting
to calculate the longitudinal structure function $F_L = F_2 - 2 x
F_1$, extracting its explicit dependence on both $\Lambda^2/q^2$ and
$1/N^2$, obtaining
\begin{eqnarray}
F_L &=& F_2 - 2 \, x \, F_1 \nonumber \\
&=& f^{(0)}_2 \, \left(\frac{\Lambda^2}{q^2}\right)^{\Delta-1} +
\frac{1}{N^2} \, \left( f^{(1)}_2 - 2 \, x \, f^{(1)}_1 \right) \,
\left(\frac{\Lambda^2}{q^2}\right) + \frac{1}{N^4} \, \left(
f^{(2)}_2 - 2 \, x \, f^{(2)}_1 \right) \,
\left(\frac{\Lambda^2}{q^2}\right) + \cdot \cdot \cdot \label{FL}
\nonumber \\
&&
\end{eqnarray}
with $\Delta \geq 4$ (where $\Delta$ is the conformal dimension
associated with the incoming dilaton), in such a way that the
functions $f^{(n)}_1$ and $f^{(n)}_2$ (where $n=0, 1, \cdot \cdot
\cdot$) give the order in $1/N^2$ corresponding to the expansions of
$F_1$ and $F_2$, respectively. Notice that $F_L$ in principle
contains all the terms of the form $1/N^{2n} \times \left( f^{(n)}_2
- 2 \, x \, f^{(n)}_1 \right) \, \left(\Lambda^2/q^2\right) $, which
correspond to the exchange of $(n+1)$-intermediate states in the
FCS, {\it i.e.} corresponding to $(n+1)$-hadron final states in DIS.
From equation (\ref{FL}) we can observe several interesting aspects.
For instance, the large $N$ limit and the limit in which $q^2 \gg
\Lambda^2$ do not commute, which means that at infinite $N$ the
first term is the leading one, implying that the dominant
contribution to DIS in this limit comes from single-hadron
intermediate states in the FCS. On the other hand, if we first take
the $q^2 \gg \Lambda^2$ limit, the second term dominates (after
considering $N \gg 1$), indicating that two-particle intermediate
states give the leading contribution. Recall that this is the
so-called high energy limit. Moreover, as we will show below, the
rest of the contributions in this limit are subleading under certain
assumptions that will be discussed later. There is an explicit
tensor structure associated with each term in $F_1$ and $F_2$ in the
expansion above that we will study in this work. This allows us to
provide a strong argument in favor of the structure of the expansion
of equation (\ref{FL}).

~

The paper is organized as follows. In the rest of this Introduction
we study DIS beyond the $N \rightarrow \infty$ limit and then we
briefly comment on the operator product expansion analysis of DIS at
strong coupling. In sections 2 and 3 we perform the supergravity
calculation of diagrams with two intermediate states in a detailed
way. In Section 4 we consider some general aspects of supergravity
diagrams involving multi-particle intermediate states, which imply
$1/N^{2n}$ corrections to the FCS and DIS processes. In Section 5 we
present the discussion and conclusions. Some details of our
calculations are presented in Appendices A and B.

\subsection{Two-particle intermediate states in FCS}

The aim of the present work is to study the leading $1/N^2$
corrections to the scalar glueball structure functions in the
strongly coupled regime of the gauge theory. Therefore, it is
important to understand how this affects the calculation of the
supergravity amplitude.

Within the AdS/CFT correspondence, the regime where classical
supergravity is an accurate description of the boundary field theory
is the planar limit, where the 't Hooft coupling $\lambda=g_{YM}^2
N$ is kept fixed and large with the condition $1 \ll \lambda \ll N$.
It is possible to go beyond this approximation in two directions
given by two series expansions, one in powers of $1/\sqrt{\lambda}$,
while the other one is the $1/N$ expansion, which for adjoint fields
leads to a $1/N^2$ expansion. From the dual string theory point of
view the strong coupling expansion ($1/\sqrt{\lambda}$) and the
$1/N^2$ one correspond to the $\alpha'$ expansion and the genus
expansion ({\it i.e.} the string coupling $g_s$ expansion),
respectively. In the low energy limit of type IIB superstring theory
the genus expansion of type IIB string theory becomes a loop Feynman
diagram expansion in type IIB supergravity, and this is the one that
we study.

In the $N \rightarrow \infty$ limit only tree-level diagrams must be
included. In fact, since we consider the low energy limit of type
IIB superstring theory in the large $N$ limit we use type IIB
supergravity at tree level. In the holographic dual calculation of
DIS for scalar glueballs\footnote{We refer to this process as the
holographic dual of DIS, but as we have seen formally this is the
holographic dual description of FCS.} we are dealing with a $2
\rightarrow 2$ scattering process between two gravitons and two
dilatons. Therefore, in this limit we only need to study the Witten
diagrams corresponding to $s$-, $t$- and
$u$-channels\footnote{Recall that $s$, $t$, and $u$ are the
Mandelstam variables.}, together with diagrams with four-point
interaction vertices. In this case and also for other types of
hadrons such as (holographic) mesons, it can be shown that the
$s$-channel diagram is the relevant one when the center-of-mass
square energy $s = -(P+q)^2$ is not large enough in order to produce
excited string states in the holographic dual process
\cite{Polchinski:2002jw,Koile:2011aa,Koile:2013hba}. However, at
high energy the $t$-channel graviton exchange dominates the dynamics
of the process \cite{Polchinski:2002jw,Koile:2014vca}. Thus,
different regimes can be investigated in different ways according to
the value of the Bjorken parameter: supergravity gives the full
picture provided that $1/\sqrt{\lambda} \ll x < 1$, however when $x
\ll 1/\sqrt{\lambda}$ it is necessary to consider string theory.

\vspace{0.4cm}

Let us consider type IIB supergravity. In the Einstein frame its
action is given by
\be
S_{IIB}^{SUGRA} = - \frac{1}{2 \kappa_{10}^2} \, \int d^{10}x \,
\sqrt{|\det g|} \, \bigg[ {\cal {R}}_{10} - \frac{1}{2} \,
(\partial\phi)^2 -  \frac{1}{2} \, e^{2 \phi} \, (\partial {\cal
{C}})^2 - \frac{1}{4 \cdot 5!} \, (F_5)^2 \bigg] \, ,
\ee
where $\phi$ is the dilaton, ${\cal {C}}$ is the Ramond-Ramond axion
field and $F_5$ is the five-form field strength. This action must be
supplemented with the self-dual condition for the five-form field
strength.

An exact solution is the AdS$_5 \times S^5$ background metric
\be
ds^2 = \frac{r^2}{R^2} \, \eta_{\mu\nu} \, dx^\mu dx^\nu +
\frac{R^2}{r^2} \, dr^2 + R^2 \, d\Omega^2 \, , \label{adsmetricr}
\ee
where $R^4=4 \pi g_s N \alpha'^2$. In order to fix notation indices
$M, N=0, 1, \cdot \cdot \cdot, 9$ are on AdS$_5 \times S^5$, Greek
indices $\mu, \nu=0, 1, \cdot \cdot \cdot, 3$ and Latin indices $m,
n=0, 1, \cdot \cdot \cdot, 4$ are on AdS$_5$, while Latin indices
$a, b=5, 6, \cdot \cdot \cdot, 9$ are on $S^5$.

Now we describe how to perform the $1/N$-power counting in type IIB
supergravity Feynman diagrams. For that we must carry out the
dimensional reduction of type IIB supergravity on $S^5$ (see for
instance \cite{D'Hoker:1999pj} and also
\cite{Koile:2013hba,Liu:1999kg,Freedman:1998tz,D'Hoker:1999jp}). The
resulting reduced action can be written in terms of the
five-dimensional dilaton $\phi_5(x)$ as
\be
S_{5d}^{SUGRA} = - \frac{1}{2 \kappa_{5}^2} \, \int d^{5}x \,
\sqrt{|\det g_5|} \, \bigg[ {\cal {R}}_{5} - \frac{1}{2} \,
(\partial\phi_5)^2 + \dots \bigg] \, .
\ee
In this action dots indicate other terms which are not relevant in
our calculation, since we only consider the $1/N^2$ series
expansion. The constant $\kappa_5$ is given by
\be
\frac{1}{2 \kappa_5^2} = \frac{N^2}{8 \pi^2 R^3} \, .
\ee
Next, we define the canonically normalized fields, namely: we
rescale the five-dimensional dilaton as $\tilde\phi_5 \equiv N
\phi_5$, and also we do this for the graviton. Thus, by plugging the
canonically normalized fields into $S_{5d}^{SUGRA}$ we obtain the
$1/N$ dependence of the three-point interaction vertices and
$1/N^2 $ dependence of the four-point ones. With them we can
construct the Witten diagrams with the corresponding $1/N$-power
counting.

\vspace{0.4cm}

In order to obtain the one-particle exchange contribution to the
hadronic tensor and the structure functions $F_1$ and $F_2$, it is
necessary to calculate the imaginary part of the amplitude
associated with the $s$-channel interaction between two dilatons
$\phi_{in}$ and $\phi_{out}$, and two metric perturbations
(gravitons) of the form $h_{m a} = A_m v_a$. In this notation, $A_m$
represents a $U(1)$ gauge field in AdS$_5$ and $v_a$ is a Killing
vector of the five-sphere. The only way that this can occur within
type IIB supergravity is through the exchange of an intermediate
dilaton $\phi_X$ state. The interaction action directly derived from
type IIB supergravity is given by
\begin{equation}
S_{A\phi\phi} = \frac{1}{2 \kappa_{10}^2} \, i {\cal {Q}} \, \int
d^{10}x \, \sqrt{-g} \, A^m \, (\phi_{in}^* \,
\partial_m \phi_X - \phi_X^* \,
\partial_m \phi_{in})\, . \label{Aphiphi}
\end{equation}
${\cal {Q}}$ is the $U(1)$ charge of the scalar field, $v^a
\partial_a Y(\Omega) = i {\cal {Q}} Y(\Omega)$, where $Y(\Omega)$
represents an spherical harmonics on $S^5$. The corresponding
five-dimensional reduced interaction action is obtained by
integrating over $S^5$. Taking into account the dilaton rescaling
and also ${\tilde{A}}_5^m \equiv N A_5^m$, it leads to
\begin{equation}
S^{{\tilde A} \tilde\phi \tilde\phi}_{5d} = \frac{1}{8 \pi^2 N R^3}
\, i {\cal {Q}} \, \int d^{5}x \, \sqrt{|\det g_5|} \,
{\tilde{A}}_5^m \, (\tilde\phi_{in, 5}^* \,
\partial_m \tilde\phi_{X, 5} - \tilde\phi_{X, 5}^* \,
\partial_m \tilde\phi_{in, 5})\, ,
\end{equation}
which gives a factor $1/N$ for each ${\tilde A} \tilde\phi
\tilde\phi$ vertex. Thus, the tree-level diagram has an overall
factor $1/N^2$, which will also be present in all the rest of loop
diagrams. Since we are interested in the relative power counting
between different terms in the $1/N$ expansion we will ignore the
overall factor. Henceforth, we will omit the tilde on the fields.

The functional form of the non-normalizable gauge field $A_m$
dictates that the interaction must occur at $r_{int} \sim q R^2 \gg
r_0 = \Lambda R^2$. Then, as explained, the imaginary part of the
FCS amplitude is obtained by using the optical theorem, cutting the
diagram in the only possible way as shown in figure 2. Thus, one has
to evaluate the on-shell action $S_{A\phi\phi}$ and sum over all
possible intermediate states. Note that the restriction to the
$s$-channel diagram implies that the photon strikes the whole
hadron, and in the case of a scalar object this leads to $F_1=0$.
This calculation is shown schematically in figure 2
\footnote{Details of the calculation are presented in
\cite{Polchinski:2002jw, Koile:2011aa, Koile:2013hba}.}. The final
result for a scalar glueball state with scaling dimension $\Delta$
has been obtained in \cite{Polchinski:2002jw} leading to
\begin{equation}
F_2(x,q^2) = \pi \, A_0 \, {\cal {Q}}^2
\left(\frac{\Lambda^2}{q^2}\right)^{\Delta-1}x^{\Delta+1}(1-x)^{\Delta-2}
\, ,
\end{equation}
where $A_0 = 2^\Delta \pi |c_{in}|^2 |c_X|^2 \Gamma(\Delta)^2$, with
$c_{in}$ and $c_X$ being dimensionless constants.

Next, we want to calculate the first correction to these structure
functions, {\it i.e.} the leading order $1/N^2$ contribution. This
means that we have to take into account all possible type IIB
supergravity one-loop corrections to the $s$-channel diagram of
figure 2. In order to illustrate it, in figure 3 a few examples of
the one-loop diagrams which can be constructed with the available
interactions (that will be described in section 2) are shown. From
the $S^{{\tilde A} \tilde\phi \tilde\phi}_{5d}$ action it is easy to
see that since a one-loop Feynman diagram has two more vertices of
the type ${\tilde A} \tilde\phi \tilde\phi$ (or a quartic vertex) in
comparison with the tree-level Feynman diagram, then there is an
additional overall factor $1/N^2$.
\begin{figure}[h]
\begin{center}
\includegraphics[scale=0.75]{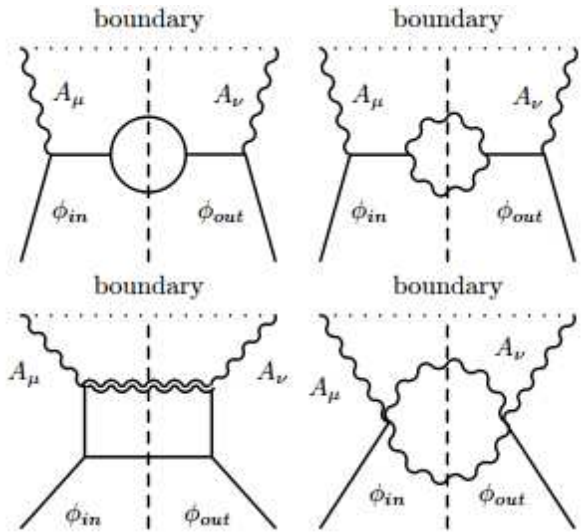}
\caption{\small Some type IIB supergravity Feynman diagrams that
could contribute to the one-loop correction to the FCS. Solid, wavy
and double wavy lines correspond to scalar, vector and tensor
perturbations from type IIB supergravity, respectively.}
\end{center}
\label{1loop}
\end{figure}
Notice that the cuts (vertical dashed lines) in these diagrams are
only schematic: the actual computation of the imaginary part of FCS
requires to square the sum of all possible supergravity Feynman
diagrams having two intermediate on-shell states. Therefore, one
also must consider the \textit{crossed} terms. This calculation is
difficult, specially in an AdS$_5$ background. A recent paper by Gao
and Mou \cite{Gao:2014nwa} has done a first step to attempt to
address these $1/N^2$ corrections. However, their calculations are
carried out in the context of an effective model given by a
scalar-vector Lagrangian, which has a very small number of modes and
interactions among them in comparison with the actual possible field
fluctuations of type IIB supergravity.

In the present work we will study this problem using the full
spectrum of particles and interactions from type IIB supergravity on
AdS$_5 \times S^5$ and show that the $\Lambda^2/q^2 \rightarrow 0$
limit renders important simplifications, leading to only one
dominant diagram. The actual scattering amplitude is difficult to
calculate, however our final formula will allow us to draw some
conclusions about the physics of this process. We will also comment
on what these observations imply on the field theory side.

\subsection{Operator product expansion analysis of DIS}

In this subsection we describe the OPE analysis of the DIS process
in the strong coupling regime of gauge theories. We follow the
analysis by Polchinski and Strassler \cite{Polchinski:2002jw}, and
describe it here since it will be relevant for the results of the
present work. Let us consider the DIS process from the quantum field
theory point of view. It is possible to perform this kind of
analysis in any SYM theory like ${\cal{N}}=4$ SYM whose conformal
invariance is broken by an IR cutoff $\Lambda$. The important point
is to have an IR confining gauge theory. It is interesting to
consider the moments of the structure functions involved in the
hadronic tensor, generically defined as
\begin{equation}
M_n^{(s)}(q^2) = \int_0^1 dx\, x^{n-1} \, F^{(s)}(x,q^2) \, .
\end{equation}
These moments can be studied in terms of the OPE of two
electromagnetic currents inside the hadron $J(0)\tilde{J}(q)$, whose
matrix element defines the hadronic tensor. In
\cite{Polchinski:2002jw} it was found that
\begin{eqnarray}
M_n^{(s)}(q^2) \approx &&\frac{1}{4}\sum_j C_{n,j}^{(s)} A_{n,j}
\left(\frac{\Lambda^2}{q^2}\right)^{\frac{1}{2}\tau_{n,j}-1} +
\frac{1}{4}\sum_{Q_p=Q} C_{n,p}^{(s)}
A_{n,p}\left(\frac{\Lambda^2}{q^2}\right)^{\tau_{p}-1}
\label{OPE}\\
&&+ \frac{1}{4N^2}\sum_{Q_p\neq Q} C_{n,p}^{(s)}
a_{n,p}\left(\frac{\Lambda^2}{q^2}\right)^{\tau_{p}-1}, \nn
\label{moments}
\end{eqnarray}
where $C^{(s)}$'s are numerical coefficients, $A$'s stand for matrix
elements of the corresponding operators and $A_{n,p} = a_{n,p}
N^{-2}$, while $\tau$'s account for their twist, given in terms of
the conformal dimension $\Delta$, the anomalous dimension $\gamma$
and the spin $s$
\begin{equation}
\tau = \Delta + \gamma - s \, .
\end{equation}
Equation (\ref{OPE}) contains very important physical information.
In the DIS regime the square momentum of the virtual photon is very
large with respect to the IR confining scale\footnote{ The limit
$q^2\rightarrow \infty$ means $q^2 \gg \Lambda^2$.} $\Lambda$ ,
therefore the lowest twist operators dominate since their
contribution are less suppressed. The first term corresponds to the
contribution to the current-current OPE coming from single-trace SYM
operators $T_{n,j}$. Using the normalization of the local operators
in such a way that they create hadrons at order $N^0$, the OPE
coefficients and matrix elements have the following behavior
\begin{eqnarray}
C_{n,j}^{(s)} &=& \langle Q,P|J\, J\, T_{n,j} |Q,P \rangle \sim N^{-1} \,  , \nn \\
A_{n,j} &=& \langle Q,P|T_{n,j} |Q,P \rangle \sim N^{-1} \, . \nn
\end{eqnarray}
The anomalous dimension of the $T_{n,j}$'s is of order $\gamma \sim
\lambda^{1/4}$. Twist-two single-trace operators give the dominant
contribution at weak 't Hooft coupling. However, when the coupling
becomes large this is no longer the case. The second and third terms
are associated with certain double-trace operators $T_p^{\dagger}
(\partial)^r T_p$ built from the so-called protected operators
$T_p$. The conformal dimension of the protected operators has small
or null corrections. Therefore, protected double-trace operators
have the lowest twist and dominate the OPE when $\lambda$ is
sufficiently large. In addition, it can be seen that among these
operators there are two possibilities \cite{Polchinski:2002jw},
namely:
\begin{eqnarray}
A_{n,p} &=& \langle Q,P|T_p^{\dagger} (\partial)^r T_p |Q,P \rangle
\sim N^{0} \,\,\,\, \textrm{if} \,\,\,\, \langle Q,P|T_p|0\rangle \neq 0 \label{A2} \, , \\
A_{n,p} &=& \langle Q,P|T_p^{\dagger} (\partial)^r T_p |Q,P \rangle
\sim N^{-2} \,\,\,\, \textrm{if} \,\,\,\, \langle Q,P|T_p|0\rangle =
0 \label{A3} \, .
\end{eqnarray}
The second one corresponds to the third term in equation
(\ref{OPE}). Obviously, at $N\rightarrow \infty$ this term is
negligible and the OPE is dominated by the second term, and we can
see from equation (\ref{A2}) that it describes a regime where hadron
production is turned off. However, at finite $N$ the hadron number
is not conserved, and the third term becomes important. In fact, we
will see that this is the leading contribution since in this case
the lower twist contributions can come from the created hadrons
instead of the initial one. This is interpreted as a situation where
the virtual photon strikes a pion in the hadron cloud that surrounds
the incoming hadron.

On the one hand, the perturbative gauge theory analysis allows us to
study the weak coupling regime. On the other hand, string theory and
supergravity help us to study the strongly coupled regime. Let us
focus on the case when the Bjorken parameter is within the range
$0.1<x<1$, where the bulk physics can be accurately described by
type IIB supergravity. Then, the process can be understood in the
following way: the current operator insertion on the boundary theory
generates a non-normalizable vector fluctuation of the metric (as
seen in the five-dimensional reduction of type IIB supergravity)
which couples with the normalizable bulk modes corresponding to
hadronic states in the SYM theory. The leading behavior in the $1/N$
expansion was studied in \cite{Polchinski:2002jw} for the dilaton
and the dilatino and in \cite{Koile:2011aa,Koile:2013hba} for scalar
mesons and polarized vector mesons by using the optical theorem,
where the leading contribution comes from a diagram with no mixing
and with only one intermediate state. In this work, we build on the
work started in \cite{Gao:2014nwa} and focus on the finite $1/N^2$
leading contributions, by considering an external supergravity state
given by a dilaton (which is the dual supergravity field of a scalar
glueball state in the gauge field theory), by allowing a second
intermediate state and explicitly obtain the resulting structure
functions. This is equivalent to study the one-loop contribution to
the supergravity interaction with two external gravi-photon states
and two external dilaton states.

In principle, we would have to calculate every possible contribution
coming from a Cu\-tovs\-ky-cut diagram allowed by type IIB
supergravity, including all the Kaluza-Klein towers of modes from
all the fields which develop fluctuations. Among them, for example,
we can find the ones coming from the three-scalar vertex considered
in \cite{Gao:2014nwa}. This is complicated, since the geometry of
AdS$_5$ renders Bessel function solutions, and then the integral of
a generic three-particle interaction would be impossible to be
carried out analytically. However, the OPE of equation (\ref{OPE})
gives us an important insight into the physical process that we are
trying to describe.

Reference \cite{Polchinski:2002jw} shows that when the current
operator couples directly to a state of Kaluza-Klein mass $\Delta$
the resulting scattering amplitude (and the structure functions) are
proportional to $\left(\Lambda^2/q^2\right)^{\Delta-1}$. This should
hold regardless of the fact that this hadron might not be the
initial state, since it could come from a hadron splitting into two
other hadrons. This hypothesis is supported by the $1/q^2$-power
analysis performed in \cite{Gao:2014nwa}: by looking at the $s$-,
$t$- and $u$-channel (one-loop) diagrams with scalars, we expect
that the less suppressed contribution would come from the
$t$-channel where the mode with the lowest Kaluza-Klein mass is
exchanged (corresponding to the lowest-twist coupling). This is
exactly what happens. In fact, the interaction terms present in the
action imply that this is the diagram which dominates the full
amplitude at strong coupling and finite $N$. The rest of diagrams
are suppressed by higher powers of $1/q^2$. This was anticipated in
reference \cite{Polchinski:2002jw}.

In the rest of this paper we will obtain this particular leading
amplitude and calculate the structure functions with $1/N^2$
corrections.

\section{Supergravity calculation of diagrams with two intermediate states}

\subsection{The background and its $S^5$-reduced spectrum}

The background used in this work is a deformation of type IIB
supergravity on the AdS$_5\times S^5$ of radius $R$, which can be written as
\begin{equation}
ds^2 = \frac{R^2}{z^2}\left(\eta_{\mu\nu}dx^\mu dx^\nu + dz^2\right)
+ R^2 d\Omega_5^2 \, .
\end{equation}
When $z=R^2/r$ this becomes the metric (\ref{adsmetricr}). In this
coordinate system, the conformal boundary of the AdS space is
located at $z=0$, or equivalently at $r\rightarrow \infty$ in
equation (\ref{adsmetricr}). By introducing a cutoff $r_0$ it
corresponds to an IR confinement scale of the boundary gauge theory
$\Lambda=r_0/R^2 = z_0^{-1}$. Recall that the self-dual five-form
field strength $F_5$ has $N$ units of flux through the five-sphere.
At low energy with respect to $1/\sqrt{\alpha'}$ the spectrum of
fluctuations of type IIB supergravity is similar to the one
described in \cite{Kim}.

Now, let us briefly review how the full spectrum of bosonic
fluctuations around the AdS$_5\times S^5$ background is calculated.
The relevant fields contained in the bosonic part of the action are
the metric $G_{MN}=g_{MN}+h_{MN}$, the complex scalar $\phi$ and the
RR four-form $A_4$ ($F_5=dA_4$ in this case). The non-zero
components of $F_5$ with no fluctuations are
\begin{equation}
F_{mnopq} = \frac{1}{R} \epsilon_{mnopq} \ , \ F_{abcde} =
\frac{1}{R} \epsilon_{abcde} \, ,
\end{equation}
where the $\epsilon$ stands for the Levi-Civita pseudo-tensor
density. Recall that the zeroth order metric $G$ and $F_5$ are
non-vanishing. If we want to study the corresponding fluctuations we
need to work out the equations of motion at quadratic order. One
starts from the expansion on $S^5$, leading to the usual
Kaluza-Klein decomposition of the fields in a basis of spherical
harmonics. This includes scalar, vector and tensor (symmetric or
antisymmetric) spherical harmonics\footnote{Details on the
definition and properties of these objects are given in Appendix B
of \cite{Lee:1998bxa}. In what follows, parentheses between two
indices mean interchange symmetry with the trace removed, while
brackets mean antisymmetry.} that we denote as $Y^l(\Omega)$, $Y^l_a
(\Omega)$, $Y^l_{(a,b)}(\Omega)$, $Y^l_{[a,b]}(\Omega)$,
respectively. These are all eigenfunctions of the angular
Laplacian\footnote{Notice that we denote the AdS Laplacian by
$\Box$.} $\nabla^2$ such that
\begin{equation}
\nabla^2 Y^l(\Omega) = -\frac{1}{R^2}k (k+4)Y^l(\Omega) \, ,
\end{equation}
for some integer $k$. By separating the different components of the
metric as
\begin{eqnarray}
G_{mn} = g^{(AdS)}_{mn} + \tilde{h}_{mn} \, , \ && \ \tilde{h}_{mn}=
h_{mn}-\frac{1}{3}g_{mn}^{(AdS)}h^{a}_{\phantom{ }a} \, ,
\label{DescG} \\
G_{ma} = h_{ma} \, , \ && \ G_{ab} = g_{ab}^{S^5} + h_{ab} \, ,
\nonumber
\end{eqnarray}
and by fixing the De Donder-type gauge conditions $D^{a}h_{(ab)} =
0$ and $D^{a}h_{am}=0$, we have
\begin{eqnarray}
h_{m n}(y, \Omega) = \sum_{l} H_{m n}^l(y) \, Y^l(\Omega) \, , \ &&
\
h_{m a}(y, \Omega) = \sum_{l} A_{m}^l(y) \, Y_{a}^l(\Omega) \, ,\nonumber \\
h_{(ab)}(y, \Omega) = \sum_{l} \phi^{l}(y) \, Y^l_{(a b)}(\Omega) \,
, \ && \ h_{\,a}^{a}(y, \Omega) = \sum_{l} \pi^l(y) \, Y^l(\Omega)
\, , \nonumber
\end{eqnarray}
where $y$ denotes coordinates on AdS$_5$ while $\Omega$ are the five
angular coordinates on $S^5$. The expansion behaves similarly for
the other fields. For instance, one important part of the $A_4$
fluctuations is
\begin{equation}
a_{mabc}(y, \Omega) = \sum_{l} a^l_{m}(y) \,
\epsilon_{abc}^{\phantom{\alpha \beta \gamma}de} \, D_{d} \,
Y_{e}^l(\Omega) \, . \label{DescA4}
\end{equation}
This expansion simplifies considerably the linearized equations of
motion. Still, some algebra is needed in order to diagonalize them,
and finally a set of different Kaluza-Klein towers of particles,
each one with its Kaluza-Klein mass formula, is obtained. From the
combination of the metric expansion with some of the terms coming
from $A_4$ there are three scalar particles, two vectors and one
tensor. Their equations of motion, Kaluza-Klein masses and other
properties are listed in table 1 \footnote{More complete tables
which include all the bosons and fermions can be found in \cite{Kim}
and in the review article \cite{D'Hoker:2002aw}.}. Note that the
massless state of the $h_{(mn)}$ tower corresponds to the AdS$_5$
graviton. {\renewcommand{\arraystretch}{1.4}
\begin{table}
\begin{center}
\centering
\begin{tabular}{|c|c|c|c|c|c|c|}
\hline
Field & Spin & Build from & $m^2(k)$ & $\Delta(k)$ & ${\cal{O}}_{QFT}$ & $SU(4)_R$ \\
\hline
$\phi$ & $(0,0)$  & $\phi$ & $k(k+4) \,,\, k\geq0$ & $k+4$ & Tr$\left(F^2 X^k\right)$ & $(0,k,0)$ \\
\hline
$s$ & $(0,0)$  & $h^a_{\phantom{a} a}\ , \ a_{abcd}$ & $k(k-4) \,,\, k\geq2$ & $k$ & Tr$\left(X^k\right)$  & $(0,k,0)$ \\
 \hline
$t$ & $(0,0)$  & $h^a_{\phantom{a} a}\ , \ a_{abcd}$ & $(k+4)(k+8)
\,,\, k\geq0$ & $k+8$ &
Tr$\left(F^2 \tilde{F}^2 X^k \right)$  & $(0,k,0)$ \\
 \hline
$\Omega$ & $(0,0)$ & $h_{(ab)}$ & $k(k+4) \,,\, k\geq2$ & $k+4$ &
Tr$\left(\lambda \lambda
\overline{\lambda}\overline{\lambda}X^k\right)$
 & $(2,k-2,2)$ \\
 \hline
$A_m$ & $(\frac{1}{2},\frac{1}{2})$ & $h_{m a}\ , \ a_{m abc}$ &
$(k-1)(k+1) \,,\, k\geq1$ & $k+3$ &
Tr$\left(\lambda \overline{\lambda} X^k\right)$  & $(1,k-1,1)$ \\
 \hline
$B_m$ & $(\frac{1}{2},\frac{1}{2})$ & $h_{m a}\ , \ a_{m abc}$ &
$(k+3)(k+5) \,,\, k\geq1$ & $k+7$ &
Tr$\left(F\tilde{F} \lambda \overline{\lambda}X^k\right)$  & $(1,k-1,1)$ \\
 \hline
$h_{(mn)}$ & $(1,1)$ & $h_{(mn)}$ & $k(k+4) \,,\, k\geq0 $ & $k+4$ &
Tr$\left(F\tilde{F}X^k\right)$  & $(0,k,0)$  \\
\hline
\end{tabular}
\caption{\small Some features of type IIB supergravity fluctuations
in the AdS$_5\times S^5$ background which are relevant to this work.
The integer $k$ indicates the $SO(6)\sim SU(4)_R$ irrep and defines
the corresponding Kaluza-Klein mass. Also, the operator that creates
the boundary Fock-space state corresponding to each normalizable
fluctuation is shown. The relation between the scaling dimension
$\Delta$ and $k$ is shown.}
\end{center}
\end{table}
We need the solutions to these equations. These are shown in
Appendix A. All the normalizable bosonic modes have similar form:
the modes carrying a given four-dimensional momentum $p^{\mu}$ turn
out to be of the form \footnote{The angular dependence is only
written generically.}
\begin{equation}
\Phi_{m_1\dots} \sim \epsilon_{m_1\dots} e^{ip\cdot x}z^{\alpha}
J_{\Delta(k)-2}(p z) Y^{l(k)} (\Omega)\,
\end{equation}
for some power $\alpha$ and polarization $\epsilon_{m_1\dots}$. The
main difference between the spectrum of our confining background and
the one from \cite{Kim} for AdS$_5\times S^5$ comes from the
inclusion of the cutoff $z_0$. This imposes a restriction analogous
to the one for \textit{modes in a box} \cite{Polchinski:2002jw}
which means that $p$ is restricted to be one of the infinite but
discrete set of numbers such that $J_{\Delta(k)-2}(p z_0)=0$
\footnote{Recall that $p\equiv \sqrt{\eta_{\mu\nu}p^\mu p^\nu}$. We
call this the AdS mass as opposed to the Kaluza-Klein mass.}.
Canonical normalization for the scalar states as defined in
\cite{Polchinski:2001tt} is discussed in Appendix A.

\subsection{Selection rules for the interactions}

The different scalar, vector and tensor fields we studied in the
previous section can interact with each other in complicated ways.
These interactions can be directly obtained from the type IIB
supergravity action by performing the expansion of the fields in
terms of spherical harmonics on $S^5$. The relevant vertices will be
explicitly derived in the next section. However, besides the
appearance of these vertices in the action it is important to
consider the selection rules coming from the fact that these
\textit{particles} belong to representations of the isometry group
$SO(6)\sim SU(4)$. The lowest dimensional representations in which
these fields are found can be viewed in \cite{Kim,D'Hoker:2002aw}.

The selection rules can be written in terms of the Clebsh-Gordon
coefficients of the tensor product decomposition in irreducible
representations (irreps) of $SU(4)$ given in the notation of table 1
by
\begin{eqnarray}
(0,k_1,0)\otimes (0,k_2,0) &=& \bigoplus_{i=0}^{k_2} \bigoplus_{j=0}^{k_2-i}(j,k_1+k_2-2i-2j,j) \ , \ k_2 \leq k_1, 
%
\end{eqnarray}
and similarly for the product $(0,k_1,0)\otimes (1,k_2,1)$.
Physically, a null coefficient implies that in a scattering process
where the two initial states belong to the first two irreps, a
particle belonging to the third irrep cannot be among the final
states. Together with the reduction of the ten-dimensional action to
the five-dimensional effective one, this tells us which are the
indices of the Bessel functions that can be present in the
interactions when calculating the amplitudes involved in the dual
DIS process. In terms of our solutions, these coefficients are given
by angular integrals of combinations of the different spherical
harmonics over the $S^5$ coordinates \cite{D'Hoker:1999ea,
Arutyunov:1999en}:
\begin{eqnarray}
a_{123} &=& a(k_1,k_2,k_3) = \int_{S^5} d\Omega_5 \, Y^{k_1} Y^{k_2} Y^{k_3} \, ,  \\
b_{123} &=& b(k_1,k_2,k_3) = \int_{S^5} d\Omega_5 \, Y_a^{k_3} D^a
Y^{k_2}
Y^{k_3}  \, , \\
c_{123} &=& c(k_1,k_2,k_3) = \int_{S^5} d\Omega_5 \, D^a Y^{k_1} D^b
Y^{k_2} Y_{(a b)}^{k_3} \, .
\end{eqnarray}
The first integral appears when studying an interaction between
scalars like $s$, $t$ or $\phi$, or tensor fields in the $(0,k_i,0)$
representations. The second one involves two scalars and one vector.
These two will appear in our calculations. The third one is written
for completeness and has two scalars and one $\Omega$ field (see
table 1). These factors are present in the coupling constants of the
interaction vertices.

The relevant selection rules for the diagrams that we will consider
are the following ones\footnote{Recall that we are omitting some of
the possible outgoing particles because they are not relevant in the
process we consider. Explicit examples of these selection rules can
be checked at the web page in ref. \cite{repSU4}. Note that in this
reference the notation is slightly different.}:
\begin{enumerate}
\item When two scalars in the $(0,k_1,0)$ and $(0,k_2,0)$ representations
are involved in a three-particle interaction, the relevant outgoing
particles can be
\begin{itemize}
\item $s$, $t$, $\phi$ or $h$ particles in the $(0,k_3^{(1)},0)$ rep. with $|k_1-k_2|\leq k_3^{(1)} \leq k_1+k_2$,
\item vector particles in the $(1,k_3^{(2)}-1,1)$ rep. with $|k_1-k_2|+1\leq k_3^{(2)} \leq k_1+k_2-1$,
\item$ \Omega$-scalars belonging to the $(2,k_3^{(3)}-2,2)$ rep. with $|k_1-k_2|+2\leq k_3^{(3)} \leq k_1+k_2-2$,
\end{itemize}
where all the $k_3$ indices changes in two units.
\item When a scalar particle and one vector particle belong to the
$(0,k_s,0)$ and $(1,k_v,1)$ representations interact in the same way
the possible resulting particles are
\begin{itemize}
\item $s$, $t$, $\phi$ or $h$ particles in the $(0,k_3^{(1)},0)$ rep. with $|k_1-k_2|+1 \leq k_3^{(1)} \leq k_1+k_2-1$,

\item vector particles in the $(1,k_3^{(2)}-1,1)$ rep. with $|k_1-k_2|\leq k_3^{(2)} \leq k_1+k_2$,

\item$ \Omega$-scalars in the $(2,k_3^{(3)}-2,2)$ rep. with $|k_1-k_2|+1\leq k_3^{(3)} \leq k_1+k_2-1$,
\end{itemize}
where all the $k_3$ change as before.
\end{enumerate}

Recall that all different integers $k$ associated with each particle
are bounded from below. In fact, the existing massless particles in
general correspond to the lowest representations, given by $k=1$ for
vectors and $k=0$ for scalars and tensors. There is an exception
given by the negative mass $s_{k=2}$ scalar. In addition, consider
the case of a massless vector excitation interacting with a given
scalar particle. The vector excitation can only belong to the
$(1,0,1)$ representation, while the scalar one is in the $(0,k,0)$
representation for some integer $k$ associated with its dimension
$\Delta$ as indicated in table 1. Then, the second selection rule
implies that if we are looking for outgoing $s$, $t$ or $\phi$
scalar particles, we can only have something belonging to the same
$(0,k,0)$ representation. Now, the vector representation we have
chosen can only correspond to the $A_m$ field that represents our
holographic photon, {\it i.e.} the graviton fluctuation coming from
the boundary. Thus, as in the $S_{A\phi\phi}$ interaction of
\cite{Polchinski:2002jw} \textit{there is no mixing} for an
$S_{ssA}$ vertex.

\subsection{Relevant vertices}

Some of the relevant interaction vertices are derived in this
section. We also need the propagators of some fields, which are
considered in Appendix A. Let us first focus on how the incoming
dilaton can interact with two other fields. We focus on the $\phi
\rightarrow s + \phi$ interaction, but other interactions may be
studied in the same way. The corresponding $S_{s\phi\phi}$ vertex
comes from the dilaton kinetic term\footnote{This kind of analysis
was performed in \cite{D'Hoker:2002aw}, where the authors describe
in detail an $S_{t \phi\phi}$ vertex.}
\begin{equation}
\int d^{10}x \sqrt{-G} \, G^{MN}\partial_{M} \phi \partial_{N} \phi
\, , \label{sdd1}
\end{equation}
once the mentioned fluctuations are worked out\footnote{Throughout
this section we set $R=1$, but we will recover it in the next
section.}. The relevant fluctuations are given in equation
(\ref{DescG}) (and indirectly in equation (\ref{DescA4})). The only
non-vanishing modes we consider are the scalar ones plus
$\tilde{h}_{(mn)}$ which cannot be completely turned off: their
fluctuations are given by \cite{D'Hoker:2002aw}
\begin{equation}
\tilde{h}^k_{(mn)} =
D_{(m}D_{n)}\left[\frac{2}{5(k+1)(k+3)}(\pi^k-30b^k)\right] \, ,
\end{equation}
with
\begin{eqnarray}
b^k &\equiv& t^k - s^k \nonumber \, , \\
\pi^k &\equiv& 10[(k+4)t^k + k s^k] \nonumber \, .
\end{eqnarray}
Then, we have
\begin{equation}
\sqrt{-G}\approx \sqrt{-g}\left(1+\frac{1}{2}h^M_{M}\right) \, ,
\,\,\, G^{MN}\approx g^{MN} - h^{MN} \, ,
\end{equation}
where the indices are lowered and raised using the background metric
$g$. By plugging these expressions into the action (\ref{sdd1}) for
the case $t^k=0$, and integrating by parts using the Kaluza-Klein
mass conditions ({\it i.e.} the equations of motion at quadratic
order), it leads to
\begin{eqnarray}
S_{s\phi\phi} &=& \frac{1}{2\kappa_5^2} \int_{AdS_5} dx^5 \,
\sqrt{g_{AdS_5}} \, a_{123} \times \nonumber \\
&& \left[\frac{2k_1^2}{k_1+1} s_1 D_m \phi_2 D^m \phi_3
- \frac{2}{k_1+1}D_m D_n s_1 D^m \phi_2 D^n \phi_3\right] \nonumber \\
&=& \frac{1}{2\kappa_5^2} \int_{AdS_5} dx^5 \, \sqrt{g_{AdS_5}}
\,\,a_{123} \, s_1 \phi_2 \phi_3 \times \\
&& \left[\frac{k_1^2}{k_1+1}(m_1^2 - m_2^2 - m_3^2) +
\frac{1}{2(k_1+1)}\left((m_2^2-m_3^2)^2 -
m_1^4\right)\right]\nonumber .
\end{eqnarray}
Notice that $\phi_i$ stands for the mode with $k = k_i$ of $\phi$
and the corresponding Kaluza-Klein mass $m_i^2=m_{\phi}^2(k_i)$. The
global $N^2$ factor has been discussed in the introduction and is
absorbed by a field redefinition, leaving canonically normalized
quadratic terms, triple interactions proportional to $N^{-1}$ and
quartic interactions proportional to $N^{-2}$. By writing the masses
in terms of the $k_i$ and defining $\Sigma =
\frac{1}{2}(k_1+k_2+k_3)$ and $\alpha_i = \Sigma - k_i$ we obtain
\begin{equation}
S_{s\phi\phi} = \frac{1}{2\kappa_5^2}\int_{AdS_5} d^5x \,
\sqrt{g_{AdS_5}} \, \lambda_{123} \, s_1 \phi_2 \phi_3 \, ,
\end{equation}
where the coupling constant is given by
\begin{equation}
\lambda_{123} = \frac{-8 \, \alpha_3 \, \alpha_2  (\alpha_1+2)
(\Sigma+2)}{k_1+1} \, a_{123} \, .
\end{equation}
The sign of the coupling is irrelevant for us since our final
amplitude will be proportional to $\lambda_{123}^2$. However, note
that $\lambda_{123}$ vanishes for $k_1=|k_2-k_3|$ (and also for
$k_1=k_2+k_3+4$), which eliminates some diagrams. In fact, for
$k_1=2$ the previous selection rules only allow $k_3 = k_2-2, k_2,
k_2+2$, therefore we are left with the $k_3=k_2$ case. This is
because there is no need to consider surface terms since all the
solutions under consideration are normalizable and vanish at the
boundary. Finally, when performing the integrals needed for the
on-shell evaluation in the AdS$_5$ coordinates we use the solutions
from Appendix A. First, the integration over $dx^0 \dots dx^3$
implies the four-momentum conservation. Second, since the
determinant behaves as $z^{-5}$ and all solutions are of the form
$z^2 J_{\Delta_i-2}(p z)$ we obtain a $z$-integral of the
form\footnote{The relation between $\Delta$ and $k$ is given in
table 1. Even if it is different for each type of particle, in
Appendix A we show that all the solutions come with some Bessel
function of index $\Delta-2$.}
\begin{equation}
\int_0^{z_0} dz \, z \, J_{\Delta_1-2}(a z) \, J_{\Delta_2-2}(b z)
\, J_{\Delta_3-2} (c z) \, ,
\end{equation}
where $a,b$ and $c$ are AdS masses. Although it is difficult to
solve this integral, we will analyze it in two different ways. On
the one hand, the largest contribution comes from the $z\sim z_0$
region, which means that for numeric purposes the Bessel functions
can be approximated by the asymptotic expression
\begin{equation}
J_{m}(z) \approx \sqrt{\frac{2}{\pi z}} \cos \left(z - \frac{m
\pi}{2} - \frac{\pi}{4}\right) \, . \label{Jasymptotic}
\end{equation}
This type of numerical analysis has shown to give interesting
results in our previous work \cite{Koile:2015qsa}. On the other
hand, we can have some intuition about the physics of the process
from the case $z_0 \rightarrow \infty$, where the integral is known
(see Appendix B). For our purposes it is useful to approximate it by
using a behavior which is easily seen from numerical integration:
the result is non-zero only when one of the AdS masses is the sum of
the other two \footnote{This was observed in \cite{Auluck:2012}.}.

Now, since in our diagram there are $s$ particles we need to know
how they interact with the massless vector perturbation $A_m$
generated by the current boundary insertion. This kind of
interactions has been studied before in order to obtain a more
complete knowledge of the five-dimensional effective action from
type IIB supergravity, and proved to be very useful to calculate
$n$-point correlation functions of chiral primary operators via the
AdS/CFT correspondence \cite{Lee:1998bxa, Lee:1999pj,
Arutyunov:1998hf, Arutyunov:1999fb}. The method used in these papers
is slightly different from the previous one\footnote{This is only
for technical reasons.}. It is based on using the equations of
motion together with the self-duality condition on $F_5$ rather than
the ten-dimensional action. The authors calculate the quadratic and
cubic corrections to these equations and obtain the interaction
terms present in the action leading to the corrections. Note that in
this context integration by parts and surface terms appear as field
redefinitions that simplify the interactions. Here, we only write
the result for the triple interaction between the $A_m$ and two $s$
scalars \cite{Lee:1999pj}:
\begin{equation}
S_{ssA} = \frac{1}{2\kappa_5^2}\int_{AdS_5} d^5x \, \sqrt{-g} \,
G_{123} A^m_1 s_2 \partial_m s_3 \, ,
\end{equation}
where the coupling constant can be written in terms of the indices
$k_1, k_2$ and $k_3$ as
\begin{equation}
G_{123} =
\frac{2^5(k_1+1)(\Sigma^2-\frac{1}{4})(\Sigma+\frac{3}{2})(\alpha_1-\frac{1}{2})}{(k_1+2)(k_2+1)(k_3+1)}\,
b_{123} \, .
\end{equation}
The conclusion is that $s$ modes interact with the gauge fields in a
similar way as dilaton perturbations. The case $k_2=k_3=2$ will be
important for us. Had we considered a complex scalar field, as we
will in the next section and as was done for the dilaton in
\cite{Polchinski:2002jw}, we would have found exactly the same type
of vertex with a gauge boson and the associated $U(1)$ current as in
equation (\ref{Aphiphi}). Note that this interaction term must come
from
\begin{equation}
\int d^{10}x \, \sqrt{-G} \, \left({\cal {R}}_{10} - \frac{1}{4
\cdot 5!}F_5^2\right) \textrm{ plus  }  F_5 = \star F_5 \, .
\end{equation}
In this case, by evaluating the vertex with the on-shell solutions and
integrating it leads to the four-dimensional momentum
conservation delta, now multiplied by a $z$ integral of the form
\begin{equation}
\int_0^{z0} dz \, z^2 K_{\Delta_1-2}(a z) J_{\Delta_2-2}(b z)
J_{\Delta_3-2}(c z) \, , \label{KJJ}
\end{equation}
as in the $N\rightarrow \infty$ case in \cite{Polchinski:2002jw}. We
will elaborate on this in the next sub-section. Note that the Bessel
function $K$ vanishes rapidly when going to the interior of AdS,
which means that in this case integrating up to $z\rightarrow
\infty$ is effectively the same as stopping the integration at
$z_0$.

For completeness let us discuss another situation: the quartic
vertex that would appear twice in a one-loop diagram like the fourth
one in figure 3 (with gauge or scalar intermediate particles). It is
obtained from the dilaton kinetic term in the action (\ref{sdd1}).
We expand the determinant and the metric in terms of the
fluctuations obtaining the following action in ten dimensions up to
an overall constant,
\begin{eqnarray}
S_{\phi \phi h h}&=&\int d^{10}x \sqrt{-g} \left( -\frac{1}{4} h
h^{M N} \partial_{M} \phi \partial_{N}\phi +\frac{1}{8} h^2
\partial^P \phi \partial_P \phi +\frac{1}{8} h^M_N h^N_M
\partial^P \phi \partial_P \phi \right. \nonumber \\
&& \ \ \ \ \ \  \ \ \ \ \ \ \ \ \ \  \ \ \ \ \ \ \ \ \ \left.
+\frac{1}{2} h^{M P} h _P^N \partial_{M} \phi
\partial_{N}\phi \right),
\end{eqnarray}
where $h$ denotes the trace $h^M_M$. The fields $h$ and $h_{M N}$
can be expanded in spherical harmonics and with the fluctuations of
the five-form field strength, we can build for example the $s$ and
$t$ scalar modes. The second term will not be considered since
vector fluctuations are absent. The other terms have two dilatons
coupled to $A_{m}$ and a fluctuation in the AdS$_5$ space. As we
will see in the next section, the normalizable mode of the incident
dilaton can be approximated by its asymptotic expansion near the
boundary since this is where the interaction takes place. Then, the
$z$-integral becomes proportional to the integral of two $J$ Bessel
functions and one $K$ Bessel function. The complete integral can be
calculated from equation (\ref{intJJK}), however we are interested
in the $q$-dependence
\begin{equation}
{\cal{M}}\propto \int dz \, z^{\Delta_1+\alpha} K_1(a z)
J_{\Delta_2-2}(b z) J_{\Delta_3-2}(c z) K_1(qz)\propto a^{-\Delta_1}
\left(\frac{b}{a}\right)^{\Delta_2}
\left(\frac{c}{a}\right)^{\Delta_3} \, ,
\end{equation}
where $\alpha$ is a constant which depends on the normalizable
solutions of the intermediate states.

Now, from dimensional analysis it is easy to see that with the
normalizations used in \cite{Polchinski:2002jw} the coupling
constants in triple scalar vertices with no derivatives have to be
proportional to $R^2$. This is important in order to obtain
dimensionless structure functions from the holographic FCS
amplitude. In fact, final results will not depend on $R$.

In addition, we would like to note that in a general one-loop
diagram one has to take into account fluctuations of all kind of
fields from type IIB supergravity, including fermions. We have not
discussed this here because in fact we will focus on one single
diagram, and the selection rules involved in this diagram (together
with consistent dimensional reduction) do not allow the appearance
of these fields.

\subsection{Classification of diagrams}

In the previous sections we have discussed some important aspects of
the particles present in the AdS$_5 \times S^5$ background with a
cutoff and their possible interactions. However, we have only
focused on some of them: triple interactions involving $s$ scalars,
dilatons and graviton fluctuations. In this section we will see why
these are all the interactions we need, and infer which diagrams
must be considered in the context of the one-loop supergravity dual
process of DIS.

As seen in the Introduction, the process under consideration is a $2
\rightarrow 2$ scattering where both the initial and final states
are two-particle states. There is a normalizable $\phi_\Delta$
dilaton fluctuation for some $\Delta$ and a non-normalizable
massless vector field $A_m$ which propagates from the boundary of
AdS$_5$ into the bulk. The dilaton is dual to the scalar glueball,
while the Abelian gauge field corresponds to the virtual photon.
Since the non-normalizable mode is given by a Bessel function of the
form $K_1(q z)$ it only lives near the boundary in the small $z$
region. In the $N\rightarrow \infty$ limit particle creation is not
allowed, and the incident holographic hadron has to {\it tunnel}
from the interior to this region in order to interact with it,
leading to a suppression of the scattering amplitude by the factor
$\left(\Lambda^2/q^2\right)^{\Delta-1}$. This can be interpreted as
the probability of the full hadron to shrink down to a size of order
$1/q$. The details of this calculation are given in Appendix A, but
the important part is that the interaction term
\begin{equation}
S_{A\phi\phi} = \int d^{10}x \sqrt{-G} A^m v^a \partial_m \phi
\partial_a \phi \, ,
\end{equation}
evaluated on-shell
gives an integral in the radial variable which takes the following
form
\begin{equation}
\int_0^{z_0} dz \, z^2 \, J_{\Delta-2}(P z) J_{\Delta-2}(s^{1/2} z)
K_1(q z) \approx 2^{\Delta-1} \Gamma(\Delta) \frac{q \,
s^{\frac{\Delta}{2}-1}}{(s+q^2)^{\Delta}} \, , \label{JJK}
\end{equation}
where
\begin{equation}
s=-(P+q)^2 \approx -q^2 - 2P\cdot q = -\frac{q^2}{x}(1-x) \, ,
\end{equation}
is the Mandelstam variable related to the center-of-mass energy in
four dimensions. The incoming momentum $P$ is not very large in
comparison to $q$ or $s^{1/2}$ and we can use the asymptotic
expression of $J_{\Delta-2}(z)\sim z^{\Delta-2}$ for small
arguments. Thus, after squaring the result of the integral according
to the optical theorem (and by considering the normalizations and
the sum over intermediate states) one finds that the imaginary part
of the amplitude written in terms of $q^2$ and $x$ has the
anticipated suppression factor, and similarly for the structure
functions. As explained in \cite{Polchinski:2002jw}, this is exactly
the suppression factor predicted by the field theory OPE as we can
see from the second term in equation (\ref{OPE}).

Now, the important point is that this analysis holds for any diagram
where a scalar field interacts with the $A_m$ coming from the
boundary. This is because as we have seen the vertex has the same
form. Beyond the $N\rightarrow \infty$ limit, one-loop diagrams with
different intermediate particles can contribute and one of these
particles scatters from the interaction with the dual virtual
photon. Since all the solutions have similar combinations of powers
and Bessel functions, in our calculations we should find integrals
like equation (\ref{JJK}) \footnote{The approximation of a small
argument of the incoming Bessel function could break down for an
intermediate particle. However, we will see that this possibility is
suppressed and henceforth we assume the validity of the result of
the integral.}. In consequence, we have found a hint about how each
diagram will be suppressed by powers of $\Lambda^2/q^2$, and shown
that it is directly related to the conformal dimension $\Delta$ of
the mode that interacts with the gauge field. This is where the
large $q^2$ limit becomes important: it classifies the different
diagrams according to their relative weight in powers of
$\Lambda^2/q^2$, and implies that there will be a dominant ({\it
i.e.} less suppressed) contribution. This is strongly supported by
the OPE formula (\ref{OPE}), since the third term gives a
contribution of the expected form, namely: it is suppressed by
$1/N^2$ and with different $\Lambda^2/q^2$ powers associated with
different operator twists which could be smaller than the one
associated with the full target hadron. For example, the
corresponding vertex of an $s$-channel diagram as in the first two
cases of figure 3 will produce a suppression similar to the
tree-level Witten diagram. However, when considering a diagram where
the incoming dilaton splits into two particles, only one of the
resulting pieces carrying some fraction of the original
four-momentum interacts with the graviton perturbation near the
boundary, leading to a suppression related to the nature of this
particle and its Kaluza-Klein mass, defined by a conformal dimension
$\Delta'$. This is consistent with the fact that in a process like
the one we are describing this intermediate particle is the only one
which has to tunnel to the small-$z$ region.

Our conclusion is the following: the dominant diagram or sum of
diagrams will be given by the ones where this role is played by the
particle or particles with the lowest possible $\Delta'$. This is
consistent with the expectations from reference
\cite{Polchinski:2002jw}. This analysis holds in more general cases
as we will see in section 4. Fortunately, in the one-loop case this
leads to only one possibility as the lowest $\Delta'=2$ dimension
can only be found at the bottom of the Kaluza-Klein tower
corresponding to the $s$ scalar particles of table 1 \footnote{This
kind of behavior was already found for different processes in
\cite{Polchinski:2001tt}. It was also suggested for this case in
\cite{Polchinski:2002jw}.}. Note that this excludes for example the
diagram with quartic vertices discussed in the previous section,
which will always be more suppressed.

There is an interesting feature that we can discuss. The $N
\rightarrow \infty$ limit leads to $F_1=0$ since the photon strikes
the entire scalar hadron. Beyond this limit, by considering DIS with
two-hadron final states it leads to a non-vanishing structure
function $F_1$. This is due to the fact that the incoming glueball
splits into two other hadrons and only one of them interacts with
$A_\mu$ near the boundary region. Therefore, there is a set of
diagrams which contribute in order that $F_1 \neq 0$, among which
there is the leading contribution.

From the detailed analysis carried out in this section and from the
vertices studied in the previous subsections, we conclude that the
leading diagram is the one shown in figure 4.
\begin{figure}[h]
\begin{center}
\includegraphics[scale=0.5]{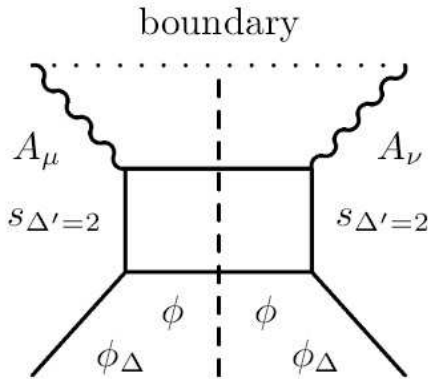}
\label{tchannel} \caption{\small Witten diagram corresponding to the
one-loop dominant contribution to FCS.}
\end{center}
\end{figure}
Although we have not written it here, we consider all the scalar
solutions to have a definite charge ${\cal {Q}}_i$ under the $U(1)$
symmetry, and assume charge conservation in each vertex. This means
that if the charge of the initial hadron is ${\cal {Q}}_1$, then the
on-shell intermediate states must have charges ${\cal {Q}}_2$ and
${\cal {Q}}_3$, such that ${\cal {Q}}_1 = {\cal {Q}}_2 + {\cal
{Q}}_3$.

We ought to say that even if all the ingredients seem to support
this conclusion, this is not a full proof. This is hard to do since
the definite integrals with three or four Bessel functions arising
from the evaluation of the amplitude and in particular from
integrations in $z$ are not known analytically in every parametric
regime (for the AdS masses) and for any combination of indices.
However, this analysis should be extensive to other theories whose
dual backgrounds are asymptotic to AdS$_5 \times S^5$. In fact, for
any asymptotically AdS$_5 \times C^5$ background, where $C^5$ stands
for some compact five-dimensional Einstein manifold, the idea would
be the same: to find the excitation with the smallest conformal
dimension and construct the diagram or diagrams where the initial
hadron produces this particle, which is the one that interacts with
the holographic virtual photon.

\section{Results for the structure functions}
\subsection{General considerations and tensor structure}

Once the leading diagram and the relevant interaction terms are
identified, we work out an expression for the imaginary part of the
scattering amplitude and extract the order $1/N^2$ contributions to
the hadronic tensor and its structure functions. The imaginary part
of $T^{\mu\nu}$ is obtained by using the optical theorem. We must
calculate the scattering amplitude for the process at the left-hand
side of the vertical cut of figure 4 with on-shell outgoing
particles, and then square the resulting amplitude and sum over all
possible intermediate states. In comparison with the $N\rightarrow
\infty$ case, there is now an off-shell state: the propagating $s$
scalar represented in this figure by a vertical line on each side of
the cut. This state is very important, since as we have seen its
conformal dimension $\Delta'=2$ ensures that we obtain the smallest
$\Lambda^2/q^2$ suppression.

This is also depicted in figure 5, where we define the momenta and
mass notation which we use in the rest of the paper.
\begin{figure}[h]
\begin{center}
\includegraphics[scale=0.5]{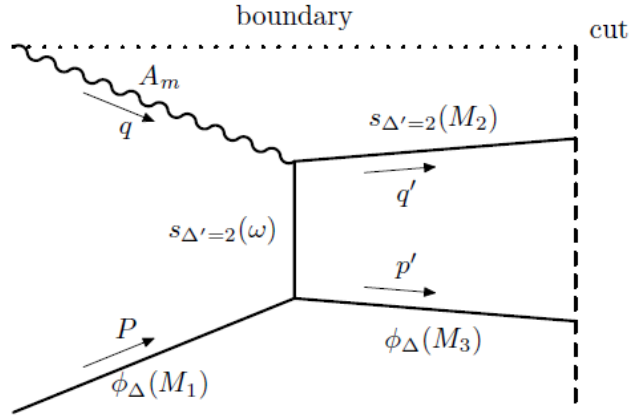}
\label{leftside} \caption{\small Feynman diagram corresponding to
the left-hand side of the cut. The field associated with each line
is explicitly presented along with the notation for the
four-dimensional momenta, conformal dimensions and AdS masses. The
solutions are described in Appendix A.}
\end{center}
\end{figure}
Notice that we use $q'^0 = \sqrt{M_2^2 + |\vec{q'}|^2}$ and $p'^0 =
\sqrt{M_3^2 + |\vec{p'}|^2}$. We will work in the center-of-mass
frame, where
\begin{equation}
P^0 = |\vec{P}|=|\vec{q}| = \frac{q}{2\sqrt{x(1-x)}}, \
|\vec{q'}|=|\vec{p'}| \textrm{ and } q^0 =
\frac{(1-2x)q}{2\sqrt{x(1-x)}} \, .
\end{equation}
Now, let us define the following vectors as
\begin{equation}
v_s^\mu = \frac{1}{q}\left(P^\mu + \frac{q^\mu}{2x}\right)\
\textrm{and} \ v_t^\mu = \frac{1}{q}\left(q'^\mu +
\frac{q^\mu}{2y'}\right) \ \textrm{with} \ y'=\frac{-q^2}{2q'\cdot
q} \, . \label{defvsvt}
\end{equation}
The auxiliary variable $y'$ can be thought of as equivalent to the
Bjorken parameter $x$ for the scattering of the $s$ scalar and the
gauge field.

In the field theory side, this will lead to the dominant
contribution of the $W^{\mu\nu}$ hadronic tensor for interactions
with two intermediate states $X_1$ and $X_2$, and we can
schematically write
\begin{eqnarray}
\textrm{Im}\left(T_2^{\mu\nu}\right) &=& \pi \sum_{X_1,X_2}
\langle P,Q|\tilde{J}^\mu (q)|X_1,X_2\rangle\langle X_1,X_2|J^\nu (0)|P,Q\rangle  \\
&=& \pi \sum_{M_2, M_3}\int \frac{d^3p'}{2E_{p'}(2\pi)^3}
\frac{d^3q'}{2E_{q'}(2\pi)^3}\langle P,Q|\tilde{J}^\mu (q)|X_1,X_2\rangle\langle X_1,X_2|J^\nu (0)|P,Q\rangle \nonumber \\
&=& 4\pi^3 \sum_{M_2,M_3} \int \frac{d^4q'}{(2\pi)^4} \delta
\left(M_2^2-q'^2\right) \delta \left(M_3^2-(P+q-q')^2\right)|\langle
P,Q|J^\nu (0)|X_1,X_2\rangle|^2,\nonumber
\end{eqnarray}
where the subindex in $T^{\mu\nu}_2$ indicates that we are
considering only processes with two-particle intermediate states,
and
\begin{equation}
n_\mu \langle P,Q|\tilde{J}^\mu (q)|X_1,X_2 \rangle = (2\pi)^4
\delta^{4}\left(P+q-p'-q'\right) \langle P,Q|n \cdot J
(0)|X_1,X_2\rangle \, ,
\end{equation}
is identified in the AdS/CFT context with what we have been calling
the amplitude on each side of the cut. Thus, we obtain the following
hadronic tensor
\begin{equation}
W_2^{\mu \nu} = \sum_{M_2,M_3} c^2 \int \frac{d^3p'}{2
E_{p'}(2\pi)^3} \frac{d^3p'}{2 E_{p'}(2\pi)^3} (2\pi)^4
\delta^4\left(P+q-p'-q'\right) v_t^{\mu}v_t^{\nu} |C_t|^2 \, ,
\end{equation}
where $c^2 \equiv c_1^2 c_2^2 c_3^2$ stands for square of the
product of the normalization constants of each on-shell field. The
complex scalar factor $C_t$ contains all the information from the
evaluation of the vertices and the propagator of the diagram as well
as from the AdS$_5$ solutions, with the exception of the phase
factors and the integrals on the $x^{\mu}$ coordinates which only
give the four-momentum conservation. By plugging the explicit
solutions and the propagator given in Appendix A, equation
(\ref{propagator}), we can schematically write
\begin{equation}
C_t(M_2, M_3, p', q') = \int dz \, dz' \left[ V_{ssA}(z) \times
V_{s\phi\phi}(z') \times G(z,z')\right] \,,
\end{equation}
where we omit the integration of the spherical harmonics on the
$S^5$ whose contribution was explained in the previous sections.

Then, from the $v_t^\mu v_t^\nu$ factor of this holographic hadronic
tensor, which is consequence of that the $t$-channel diagram gives
the leading contribution, it is easy to separate the $1/N^2$
contributions to each structure function. This is because from
equation (\ref{tensorW}) we know that \footnote{Note that $q_\mu
W^{\mu \nu} = q_\nu W^{\mu \nu} = 0$, meaning that upon contraction
with the leptonic tensor $l_{\mu \nu}$ terms with $q^{\mu}$ vanish,
thus we can ignore them.}
\begin{eqnarray}
\left(\eta_{\mu \nu}-\frac{q_\mu q_\nu}{q^2}\right) W^{\mu \nu} &=&
\eta_{\mu \nu} W^{\mu \nu} = 3 q^2 F_1 + 2x v_s^2 F_2 \, , \\
&& \nonumber \\
q^2(v_s)_\mu (v_s)_\nu W^{\mu \nu} &=&P_\mu P_\nu W^{\mu \nu} = q^2
v_s^2 F_1 + 2x v_s^4 F_2 \, .
\end{eqnarray}
Thus, we obtain
\begin{eqnarray}
F_1(x,q^2) &=&  \sum_{M_2,M_3} c^2 \int \frac{d^3p'}{2
E_{p'}(2\pi)^3} \frac{d^3p'}{2 E_{p'}(2\pi)^3}  (2\pi)^4
\delta^4\left(P+q-p'-q'\right) |C_t|^2 \nonumber \\
&& \,\,\,\,\,\,\,\,\,\,\,\,\,\,\,\,\,\,\,\,\,\,\,\,\,\,\,\,
\times 2q^2 \left[v_t^2 + 4x^2 (v_s \cdot v_t)^2\right] ,\label{F1}\\
F_2(x,q^2) &=& \sum_{M_2,M_3} c^2 \int \frac{d^3p'}{2
E_{p'}(2\pi)^3} \frac{d^3p'}{2 E_{p'}(2\pi)^3}  (2\pi)^4
\delta^4\left(P+q-p'-q'\right) |C_t|^2 \nonumber \\
&& \,\,\,\,\,\,\,\,\,\,\,\,\,\,\,\,\,\,\,\,\,\,\,\,\,\,\,\, \times 4
x q^2 \left[v_t^2 + 12 x^2 (v_s \cdot v_t)^2\right] \, .\label{F2}
\end{eqnarray}
As we will see in the next section this decomposition holds in a
more general situation. Before obtaining $|C_t|^2$ we can already
see that the first two terms which contribute to the structure
functions $F_1$ and $F_2$ can be thought of as related by the
Callan-Gross relation
\begin{equation}
F^*_2(x,q^2) = 2 \, x \, F^*_1(x,q^2) \, ,
\end{equation}
where the star means that these are not the complete structure
functions but only the first term between brackets in the
corresponding leading $1/N^2$ contribution. In contrast, the second
terms in $F_1$ and $F_2$ give non-zero contributions to the
longitudinal structure function
\begin{equation}
F_L = F_2 - 2 \, x \, F_1 \, .
\end{equation}
This will be important when analyzing our results in terms of the internal
structure of the hadron. We will discuss more about this in Section 5.

\subsection{Details of the amplitude computation}

Now, let us consider some details of the calculation of the
structure functions, {\it i.e.} the computation of (\ref{F1}) and
(\ref{F2}).

There are different parts of the calculation that we have to
assembly. First, we discuss the terms that are common to both
structure functions: the momenta integrations, the sum over
intermediate states and the complex scalar $C_t$ with the
contribution of both vertices. Then, we will write the dimensionless
factors, which define each structure function in terms of the
relevant kinematic parameters. Note that at the end all $R$ factors
cancel, thus we will omit them.
\begin{itemize}

\item There is an integral over the space component of the momenta $\vec{p'}$
and $\vec{q'}$, as well as a factor associated with the
energy-momentum conservation. This can be easily rewritten in the
center-of-mass frame and by using spherical coordinates, where all
the integrals but one can be solved trivially. The remaining one is
an angular integral in the variable $\theta$, the angle between the
incoming and outgoing vector momenta $q$ and $q'$,
\begin{equation}
\int \frac{d^3q'}{(2\pi)^32E_{q'}}\int \frac{d^3p'}{(2\pi)^32E_{p'}}
(2\pi)^4 \delta^{(4)}(P+q-p'-q') (\dots) =
\frac{|\vec{p'}|}{8 \pi q}\sqrt{\frac{x}{1-x}} \int d\theta \sin
\theta (\dots) \nonumber
\end{equation}
where $|\vec{p'}|$ solves the algebraic equation
\begin{equation}
q\sqrt{\frac{1-x}{x}}=\sqrt{|\vec{p'}|^2+M_2^2} +
\sqrt{|\vec{p'}|^2+M_3^2} \, . \label{Defp}
\end{equation}
\item There is a factor $c^2$ corresponding to the product of the normalizations of all
states involved in the process given by
$c_1^2c_2^2c_3^2$. If we assume that the masses are known this is
easy to compute since in all cases the normalization integral is
dominated by the region $z\sim z_0 = \Lambda^{-1}$. The arguments of
the Bessel functions cannot be small, therefore we can use the
asymptotic expression (\ref{Jasymptotic}). In this way an on-shell
scalar field solution associated with this Bessel function $J_{\Delta-2}$
comes with a normalization constant $c_\Delta$ such that
\begin{equation}
c_\Delta^2 = \left(\frac{\sqrt{2}}{z_0|J_{\Delta-1}(kz_0)|}\right)^2 \sim
\frac{k z_0}{z_0^2} = k \Lambda \, ,
\end{equation}
up to numerical factors. In the last step we have used the fact that
since $k z_0$ is a zero of $J_{\Delta-2}$, it must be either a minimum or a
maximum of $J_{\Delta-1}$ because of the recursion relations for the
derivative of these functions.
\item There is a sum over the masses of the intermediate on-shell
states, $M_2$ and $M_3$. The masses are constrained by the energy
conservation (\ref{Defp}). Thus, we have
\begin{equation}
\sum_{M_2 M_3}\equiv \sum_{M_2=0}^{q\sqrt{\frac{1-x}{x}}}
\sum_{M_3=0}^{q\sqrt{\frac{1-x}{x}}-M_2}.
\end{equation}
\end{itemize}
The complex scalar $C_t$ contains the information of the vertices
and the propagator (\ref{propagator}), including the coupling
constants $\lambda_{123}$ and $G_{123}$ with the corresponding $k$
indices. In what follows we will collect these in a dimensionless
constant $B$ independent of $q^2$ and $x$ whose exact form is
irrelevant for our conclusions. We can take the $\omega$-integral
out in order to factorize the other integrals, obtaining
\begin{equation}
C_t= q \int d\omega  \, \frac{\omega}{\omega^2+(P-p')^2} \,
S^{(z)}_{ssA}(M_2,q,\omega) \, S^{(z')}_{s\phi \phi }(M_1, M_3 \, ,
\omega) \, ,
\end{equation}
where $S^{(z)}_{ssA}$ and $S^{(z')}_{\phi \phi s}$ are integrals
over $z$ and $z'$, respectively. We will explain briefly each term
and calculate the integral below. Thus, we have:
\begin{itemize}
\item An integral (or sum) in the variable $\omega$ of the intermediate field
$s$ and its propagator given by
\begin{equation}
\frac{\omega}{\omega^2+(P-p')^2} = \frac{\omega
}{\omega^2-M_1^2-M_3^2 +
\frac{q}{\sqrt{x(1-x)}}\left(\sqrt{|\vec{p'}|^2+M_3^2}-|\vec{p'}|\cos
\theta\right)} \, . \label{prop}
\end{equation}
\item An integral associated with the interaction between the three scalar modes
(two dilatons and the scalar $s$)
\begin{equation}
S^{(z')}_{s \phi\phi }= \int_0^{z_0} dz' z'^2 \,
J_{\Delta-2}(M_1 z')J_{0}(\omega z')J_{\Delta''-2}(M_3 z') \, ,
\end{equation}
where $\Delta$ labels the spherical harmonics corresponding to the
initial dilaton, while $\Delta''$ is associated with the intermediate
dilaton field which has mass $M_3$. The leading contribution to this
integral is given by the region $z \sim z_0 \gg 1$. Thus, we can
approximate the Bessel functions for large arguments. By considering
both approximations of the integrals and numerical integration one
finds that this integral behaves as (\ref{semiemp})
\begin{equation}
S^{(z')}_{s \phi\phi } \sim \frac{1}{\sqrt{M_1M_3}}\left[
\delta(\omega - |M_1-M_3|) \pm \delta(\omega -(M_1+M_3)) \right] \,
, \label{semi}
\end{equation}
where the dependence on $\Delta$ and $\Delta''$ is only reflected on
the $\pm$ signs in front of each term (see Appendix B). This will
allow us to perform the integral in $\omega$.

\item An integral associated with the interaction vertex
between two fields $s$ and the non-normalizable vector perturbation
$A_\mu$. By using the axial gauge the corresponding $z$-integral
becomes
\begin{equation}
S^{(z)}_{ssA} = \int_0^{z_0} dz z^2 \, K_1(q z)
J_0(\omega z)J_0(M_2 z) \, ,
\end{equation}
where the Bessel function $K_1(qz)$ quickly decreases in the bulk
which allows one to approximate the upper limit by $z\rightarrow
+\infty$. We can solve the integral using the equation
(\ref{intJJK}) with $\rho=3$, $\lambda=0$, $\mu=0$ and $\nu=1$  from
the appendix. For $\omega << q$ the expression for the Bessel
function $J_0(\omega z)$ at small arguments can be used, and this
corresponds to consider $J_0(\omega z) \sim 1$. Therefore from
equation (\ref{intJK}), we obtain
\begin{equation}
S^{(z)}_{ssA}= \frac{2q}{\left(M_2^2+q^2\right)^2} \, . \label{ssA}
\end{equation}
Notice that both for $M_2<<1$ and $M_2 \sim q$ this provides an order
$q^{-3}$ factor.
\end{itemize}
Recall that the factors which appear in the last three
items enter the definition of $C_t$, and therefore they must be
squared in order to give $|C_t|^2$ before doing the angular
integral.

Finally, we have the dimensionless factor which define the structure
function,
\begin{equation}
\left(\begin{array}{c l}
   F_1\\
    F_2\\
    F_L
\end{array} \right)  = \frac{1}{N^2} \sum_{M_1 M_2} c^2\frac{q |\vec{p'}|}{8 \pi}\sqrt{\frac{x}{1-x}} \int d\theta \sin
\theta \left(\begin{array}{c l}
    v_t^2 + 4 x^2 (v_s \cdot v_t)^2\\
    2x [v_t^2 + 12 x^2 (v_s \cdot v_t)^2]\\  16 x^3 (v_s \cdot v_t)^2
\end{array} \right) |C_t|^2 \, .
\end{equation}
Henceforth, the prefactor $1/N^2$ carries all the $N$-dependence of
the structure functions coming from the rescaled fields. Without any
approximation the dimensionless factors in the parenthesis can be
written in terms of $M_2$, $M_3$ and $\theta$ as,
\begin{eqnarray}
&& \frac{|\vec{p'}|^2}{q^2}\left[1-\cos^2\theta\right] \, , \ \
\frac{1}{(1-x)q^2}\left[ \left(q'^0+(2x-1)|\vec{p'}|\cos \theta
\right)^2\right] \, ,
\nonumber \\
&& \textnormal{and} \ \ \
\frac{1}{(1-x)q^2}\left[2x(1-x)|\vec{p'}|^2
\left(1-\cos^2\theta\right)+ \left(q'^0+(2x-1)|\vec{p'}|\cos \theta
\right)^2\right] \, ,
\end{eqnarray}
for $F_1$, $F_L$ and $F_2$ respectively.

These are all the ingredients needed for the calculation of the
structure functions.

\subsection{Angular integral and final results for the structure functions}

Among the previous discussion the most difficult part of the
calculation is the angular integral. Recall that the factor which
depends on the angle $\theta$ is $|C_t|^2$ (through $(P-p')^2$ in
the denominator of the propagator) multiplied by the combination of
$v_t^2$ and $(v_s\cdot v_t)^2$ for each of the structure functions.

The longitudinal structure function $F_L$, on which we
will focus, takes the following expression,
\begin{eqnarray}
F_L&=& \frac{1}{N^2} B^2\sum_{M_2, M_3} \Lambda ^3 M_1 M_2 M_3 \int
d\theta \sin (\theta) \frac{1}{8\pi} \sqrt{\frac{x}{1-x}} |\vec{p'}|
q^3
\left[ \frac{\left(q'^0 + (2x-1)|\vec{p'}|\cos(\theta) \right)^2 }{(1-x) q^2} \right] \nonumber \\
&& \times \left(\int d\omega \frac{\omega}{\omega^2+
\left(P-p'\right)^2 -i\epsilon} S^{(z')}_{s \phi \phi
}(\omega,M_1,M_3) S^{(z)}_{ssA}(\omega,q,M_2) \right)^2 .
\end{eqnarray}
By using equation (\ref{semi}) the integral becomes
\begin{eqnarray}
&& F_L=\frac{1}{N^2} B^2 \sum_{M_2,M_3} \Lambda ^3 M_2 \int d\theta
\sin (\theta) \frac{1}{8\pi} \sqrt{\frac{x}{1-x}} |\vec{p'}| q^3
\left[ \frac{\left(q'^0 + (2x-1)|\vec{p'}|\cos(\theta)\right)^2
}{(1-x) q^2} \right]
\nonumber \times \\
&& \left(\int d\omega \frac{\omega}{\omega^2+ \left(P-p'\right)^2
-i\epsilon}\left[ \delta(\omega - |M_1-M_3|) \pm \delta(\omega
-(M_1+M_3)) \right] S^{(z)}_{ A s s}(\omega,q,M_2)  \right)^2 .
\nonumber \\
&&
\end{eqnarray}
Let us consider the case $M_3 \ll q$ and $|\vec{p'}|\sim q$  which
leads to the leading contribution \footnote{We assume that the case
$M_3\sim q$ leads to a subleading contribution as in
\cite{Gao:2014nwa}.}. The conditions over the mass $M_1 \ll q$
implies that $\omega= |M_1\pm M_3| \ll q$ allowing one to solve
approximately the integral $S^{(z)}_{ssA}(|M_1\pm M_3|,q,M_2)$ from
equation (\ref{ssA}). In order to solve the integral in $\theta$ we
can expand the denominator in the propagator considering $M_3 \ll
|\vec{p'}|$. This restriction imposes a condition on the upper limit
of $M_2$ in the sum, since for $M_2$ close to the maximum we can see
from the definition (\ref{Defp}) that $|\vec{p'}|$ should be small
or vanishing. By expanding $p^0$ and $p'^0$, we obtain
\begin{eqnarray}
p'^0= \sqrt{|\vec{p'}|^2+M_3^2}\approx |\vec{p'}|+\frac{M_3^2}{2|\vec{p'}|}-\frac{M_3^4}{8|\vec{p'}|^3} \, ,\\
p^0= \sqrt{|\vec{p}|^2+M_1^2}\approx
|\vec{p'}|+\frac{M_1^2}{2|\vec{p}|}-\frac{M_1^4}{8|\vec{p}|^3} \, .
\end{eqnarray}
Thus, the denominator becomes
\begin{equation}
(M_1 \pm M_3)^2+(P-p')^2\approx 2 |\vec{p}| |\vec{p'}|
\left(1-\cos(\theta)\right)+\frac{|\vec{p}|}{|\vec{p'}|}
\left(M_3\pm M_1 \frac{|\vec{p'}|}{|\vec{p}|} \right)^2+{\cal
{O}}(M_1^4) \, .
\end{equation}

The largest contribution comes from the small $\theta$ region in the
term with a minus sign. This is so because for $\theta =0$, $M_3 =
\alpha M_1$ with $\alpha=|\vec{p'}|/|\vec{p}|$ is a zero of the
denominator. Therefore, we will focus on the term with the minus
sign. The possible divergence will be addressed later\footnote{In
order to calculate the integral we assume that the IR-cutoff
$\Lambda$ is small compared with the photon momentum transfer.}.
Notice that the expression above contains two Dirac deltas, but the
term we will focus on has a very simple physical interpretation: for
$M_3<M_1$ and $M_3+\omega=M_1$ it represents a process in which the
incoming hadron splits into two hadrons, each one carrying a
fraction of the incident four-momentum.

The condition $M_3 \ll |\vec{p'}|$ implies that the
dimensionless factor of $F_L$ is approximately
\begin{equation}
\left[ \frac{\left(q'^0 + (2x-1)|\vec{p'}|\cos\theta)\right)^2
}{(1-x) q^2} \right] \approx \frac{1}{x}\left[1+\sqrt{\frac{x}{1-x}}
\frac{|\vec{p'}|}{q}\left((2x-1)\cos \theta - 1\right)\right]^2 \, .
\end{equation}
Under the mentioned approximations we obtain
\begin{eqnarray}
&&F_L= \frac{1}{N^2} B^2 \sum_{M_2,M_3} \frac{\Lambda ^3 M_2
q^5}{(M_2^2+q^2)^2} \frac{|\vec{p'}|}{2\pi}
\frac{(M_1-M_3)^2}{\sqrt{x(1-x)}}  \nonumber\\
&&\times \int_0^{\pi}  \frac{d\theta \sin (\theta) \left[ 1+
\sqrt{\frac{x}{1-x}} \frac{|\vec{p'}|}{q} \left((2x-1) \cos \theta
-1\right) \right]^2} {\left[2 |\vec{p}||\vec{p'}| (1-\cos \theta)
+\frac{|\vec{p}|}{|\vec{p'}|} (M_3- M_1 \alpha)^2  \right]^2} \, .
\end{eqnarray}
The integral in $\theta$ now can be solved, and by considering the
$1/q^2$ expansion we obtain
\begin{equation}
\int_0^\pi d\theta \frac{\sin (\theta)  \left[ 1+
\sqrt{\frac{x}{1-x}} \frac{|\vec{p'}|}{q} \left((2x-1) \cos \theta
-1\right) \right]^2} {\left[2 |\vec{p}||\vec{p'}| (1-\cos \theta)
+\frac{|\vec{p}|}{|\vec{p'}|} (M_3- M_1 \alpha)^2 \right]^2} =
\frac{\left(1-2\sqrt{x(1-x)}  \frac{|\vec{p'}|}{q}
\right)^2}{2|\vec{p}|^2  (M_3- M_1 \alpha)^2}
+{\cal{O}}\left(\frac{\log q}{q^{4}}\right) \, .
\end{equation}
Then,
\begin{eqnarray}
F_L= \frac{1}{N^2} \sum_{M_2}  \frac{\Lambda ^3 M_2
q^5}{(M_2^2+q^2)^4} \frac{|\vec{p'}|}{2\pi}
\frac{\left(1-2\sqrt{x(1-x)} \frac{|\vec{p'}|}{q}
\right)^2}{2|\vec{p}|^2 \sqrt{x(1-x)}} \sum_{M_3}
\frac{(M_1-M_3)^2}{(M_3- M_1 \alpha)^2} \, .
\end{eqnarray}
From the sum over $M_3$ we keep the most important contribution,
given by the term where $M_3$ is as close as possible to $\alpha
M_1$. Recall that $M_3$ can only take a few discrete values due to
the presence of the cutoff $\Lambda$. Then, we assume a
representative value $M_3 = \alpha M_1 + \Lambda$. Thus, we take
\begin{equation}
\sum_{M_3} \frac{(M_1-M_3)^2}{(M_3- M_1 \alpha)^2}
\approx \frac{M_1^2(\alpha-1)^2}{\Lambda^2}\, .
\end{equation}
This term depends on $\alpha=\frac{|\vec{p'}|}{|\vec{p}|}$, which
implicitly depends on $M_2$ through the definition of $|\vec{p'}|$.

Then, we approximate the sum over $M_2$ by an integral, similarly to
what is done for $M_X$ in \cite{Polchinski:2002jw}. The upper limit
is given by a fraction $0<c<1$ of the center-of-mass energy $q
\sqrt{\frac{1-x}{x}}$. Notice that $c$ should be restricted by the
condition $|\vec{p'}|\gg M_3$. Then, $|\vec{p'}|$ can be written as
a function of $M_2$ and $q$ from the equation
\begin{eqnarray}
|\vec{p'}|\approx \frac{q }{2}\sqrt{\frac{1-x}{x}} -
\frac{M_2^2}{2q} \sqrt{\frac{x}{1-x}} \, .
\end{eqnarray}
Therefore, we find
\begin{eqnarray}
F_L&=& \frac{M_1^2}{N^2} B^2 \Lambda \int_0^{c q \sqrt{\frac{1-x}{x}}}
\frac{dM_2}{\Lambda} \frac{M_2 q^4}{(M_2^2+q^2)^4}
\left(1-x\left(1+\left(\frac{M_2}{q}\right)^2\right) \right)
x^4 \left(1+\left(\frac{M_1}{q}\right)^2 \right)^4 \nonumber\\
&=&\frac{1}{N^2} B^2 c^2(2-c^2) \frac{ M_1^2}{ 4 \pi q^2} x^3 (x-1)^2  \, , \label{FL2}
\end{eqnarray}
where $B$ is a dimensionless constant that contains the
corresponding coupling constants $\lambda_{123}$ and $G_{123}$ of
Section 2.3 with the $k$ indices corresponding to each particle. We
can see that $F_L$ has a maximum around $x \approx 0.6$ and vanishes
for $x=1$ as expected. Note that the $x$-dependence of this result
is independent of the value of $c$. Also, recall that the solutions
are such that the AdS masses (as $M_1$) are proportional to
$\Lambda$.

For $F_1$ the integrals in $z$, $z'$ and $\omega$ can be solved in a
similar way as for $F_L$. The main difference comes from the
dimensionless factor in the angular integral. We obtain
\begin{eqnarray}
F_1&=&\frac{1}{N^2} \sum_{M_2,M_3} \Lambda ^3 \frac{M_2
q^5}{(M_2^2+q^2)^4} \frac{1}{8\pi}
\frac{(M_1-M_3)^2\sqrt{x}}{\sqrt{(1-x)}} |\vec{p}'|
\nonumber \\
&& \times \int_0^{\pi} d\theta \sin (\theta) \frac{
\frac{|\vec{p}'|^2}{q^2}(1-\cos ^2 \theta)} {\left[2
|\vec{p}||\vec{p}'| (1-\cos \theta) +\frac{|\vec{p}|}{|\vec{p}'|}
(M_3- M_1 \alpha)^2  \right]^2} \, . \nonumber
\end{eqnarray}
The integrals over $M_2$ and $M_3$ are very complex and we can not
obtain an analytic result for $F_1$. However, if we estimate the
$q$-power counting, it turns out that the structure function $F_1$
has a $\frac{\log q}{q^4}$ dependence. Therefore, $F_1$ is
non-vanishing but subleading.

\section{Multi-particle intermediate states from type IIB supergravity}

In this section we study the situation where there are
multi-particle intermediate states in the FCS. We investigate this
by considering Witten diagrams with multi-particle intermediate
states from type IIB supergravity. The idea is to show that both the
tensor structure (and the decomposition of the scattering amplitude
in structure functions) and the $\Lambda^2/q^2$ dependence are the
same for any number of loops from the supergravity point of view. We
also give arguments to motivate the following conjecture: within the
supergravity regime, all the $n$-loop with $n\geq 1$ leading
contributions to DIS are suppressed by the same power of
$\Lambda^2/q^2$ than the $n=1$ case that we have studied in detail
in this work. We only consider Witten diagrams such that an scalar
$s$ with the smaller scaling dimension $\Delta'=2$ interacts with
the non-normalizable gauge field. We assume the separation of this
interaction region from the rest of the multi-particle exchange
process, which occurs in the IR. This is because if the first masses
are small, all the others are bounded to be of the same order due to
the form of the vertices present in the splitting of the original
hadron, which involve normalizable modes and render a $z$-integral
of the type of our $s\phi\phi$ interaction. This type of diagrams
give the most relevant contribution for the reasons explained in the
previous sections. Figure 6 schematically represents this kind of
diagrams.
\begin{figure}[h]
\begin{center}
\includegraphics[scale=0.6]{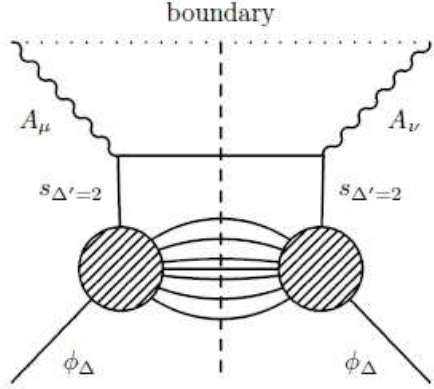}
\label{nloop} \caption{\small Relevant Witten diagrams at a
generic loop-level.}
\end{center}
\end{figure}
We can start from the most general Lorentz-tensor decomposition of
the hadronic tensor
\begin{equation}
W^{\mu\nu} = F_1(x,q^2)\left(\eta^{\mu\nu} - \frac{q^\mu
q^\nu}{q^2}\right) + F_2(x,q^2) 2 \, x \, v_s^\mu \, v_s^\nu  \, ,
\end{equation}
and the solution of the $U(1)$ gauge field which is a perturbation
of the bulk metric, induced by the current operator inserted in the
AdS boundary given by
\begin{equation}
A_\mu (x,z) = e^{iq\cdot x} \left[c_\mu qz K_1(qz) + \frac{n\cdot
q}{q^2}q_\mu\right] \ , \ c_{\mu}= n_\mu - \frac{n\cdot q}{q^2}q_\mu
\ , \ A^{\mu} = z^2 \eta^{\mu\nu}A_\nu.
\end{equation}
This solution has been obtained within the axial gauge, for which
$A_z=0$, using the boundary condition
\begin{equation}
A_\mu(x,z\rightarrow 0) = n_\mu e^{iq\cdot x} \, .
\end{equation}
The tensor structure of the amplitude is
\begin{equation}
\textrm{Im} ({\cal{A}}) \sim n_\mu n_\nu \textrm{Im} (T^{\mu\nu}) \, .
\end{equation}
The relevant interaction is the one on the vertex closer to the
boundary, given by $S_{ssA}$ or $S_{A\phi\phi}$, which appears in all the Witten
diagrams we are considering. Using the dilaton field solutions\footnote{Here we
write the steps in terms of dilatons as in the $N\rightarrow \infty$ case, but
in general they are replaced by $s$ scalars.} in the axial
gauge (see Appendix A) $\phi$, this vertex evaluated on-shell is
given by
\begin{eqnarray}
S_{A\phi\phi}|_{on-shell}&=&\int d^{10}x \, \sqrt{-g} G^{MN}\partial_M\phi\partial_N\phi \nonumber \\
&=& \int d^{10}x \, \sqrt{-g} A^mv^a\left(\partial_a\phi_1\partial_m\phi_2^{\star}
-\partial_a\phi_1\partial_m\phi_2^{\star}\right) \nn \\
&=& i{\cal{Q}} \int d^{4}y dz d\Omega_5 \sqrt{g_\Omega}\, z^{-5}
A^\mu
\left(\phi_1\partial_\mu\phi_2^{\star}-\phi_1\partial_\mu\phi_2^{\star}\right)
\nn \\
&=& i{\cal{Q}} \, \delta^{(4)}(q+p_1-p_2) \int dz d\Omega_5 \,
\sqrt{g_\Omega} \, z^{-3} \, \phi_1 \, \phi_2^{\star} \, \eta^{\mu
\nu} \, A_\mu \, \left(p_{1\nu}+p_{2\nu}\right)  \, ,
\end{eqnarray}
where $\phi_1$ is the incoming dilaton (or $s$ scalar) and $\phi_2$
is the one representing the upper intermediate state in the diagram
of figure 6. The corresponding four-momenta are $p_1$ and $p_2$,
respectively. $g_\Omega$ is the determinant of the metric of the
five-sphere with radius $R$. For $\phi$ we can chose the spherical
harmonic such that
\begin{equation}
v^a\partial_a \phi_i = c_{\phi_i} e^{ip_i\cdot x} z^2
J_{\Delta-2}(p_i z) v^a \partial_a Y(\Omega) = i {\cal{Q}}_i \phi_i.
\end{equation}
Also, charge conservation implies that ${\cal{Q}}_1 = {\cal{Q}}_2
\equiv {\cal {Q}}$. From the solution for $A_\mu$ the integral
splits into two terms: one with the Bessel function $K_1(qz)$ which
dominates in the region close to the AdS boundary, and another one
from the part of $A_\mu$ which is independent of $z$. The last one
vanishes for the reasons explained in the appendix when
$N\rightarrow \infty$, thus the tensor structure, {\it i.e.} the
factors containing $n_\mu$ is exclusively given by the square of
\begin{equation}
\eta^{\mu\nu}c_\mu\left(p_{1\nu}+p_{2\nu}\right) =
\eta^{\mu\nu}\left(n_\mu - \frac{n\cdot
q}{q^2}q_\mu\right)\left(p_{1\nu}+p_{2\nu}\right) \, .
\end{equation}
In the limit $N\rightarrow \infty$ we have
\begin{equation}
c_\mu (p_{1}^{\mu}+p_{2}^{\mu}) = c_\mu (P^\mu + (P+q)^\mu) = 2 q
(n\cdot v_s)\, ,
\end{equation}
and since $q_\mu v_s^{\mu}=0$, then $F_1=0$ and $F_2\neq0$. However,
for a one-loop amplitude we have
\begin{equation}
p_{1}^{\mu}+p_{2}^{\mu} = (q')^\mu + (P-p')^\mu = (2q'-q)^{\mu} \, .
\end{equation}
Then
\begin{equation}
\eta^{\mu\nu}\left(n_\mu - \frac{n\cdot
q}{q^2}q_\mu\right)\left(p_{1\nu}+p_{2\nu}\right) = 2 n\cdot
\left(q'+\frac{q}{2y'}\right) \equiv 2q (n\cdot v_t) \, .
\end{equation}
This tensor structure and the generic decomposition of $W^{\mu\nu}$
schematically leads to
\begin{eqnarray}
F_1 &\sim& 2q^2 \left[v_t^2 + 4x^2 (v_s\cdot v_t)^2\right] \, , \\
F_2 &\sim& 4xq^2 \left[v_t^2 + 12x^2 (v_s\cdot v_t)^2\right] \, .
\end{eqnarray}
In the above expressions we should include all the integrals which
are necessary to complete them.

It is easy to see that this analysis for one-loop Witten diagrams
holds for a generic $n$-loop diagram as schematically depicted in
figure 6. In fact, for an $n$-loop diagram the difference is that
now $p_2 = P-p'_1-\dots-p'_{n-1}$, but from momentum conservation
this is $q'-q$. The $p_i'$ are the momenta of the on-shell
intermediate particles that appear in the IR region, while $q'$ is
the momentum of the $s$ scalar after the scattering with the
non-normalizable vector. Thus, the Lorentz-tensor decomposition is
totally general, and therefore we will always have a similar
structure as the one presented in the Introduction in equation (5).
If $\theta$ is the angle between vectors $\vec{q}$ and $\vec{q}'$,
we can also say that $F_1(\theta\rightarrow 0)=0$.

Now, since the tensor structure and the most relevant vertex are the
same, we propose that the leading $q$-dependence will be the same
for all these cases. If this proposal turned out to be true there
would be an important consequence: the $1/N^{2n}$ corrections with
$n>1$ would be subleading. This would mean that once particle
creation is allowed, $N\rightarrow \infty$ and $q^2\rightarrow
\infty$ become commuting limits. In that case, the only relevant
processes in the study of DIS in the large $N$ and strong coupling
limit would be the one- and two-particle final states processes.

\section{Discussion and conclusions}

In this work we have focused on the $1/N^2$ corrections to DIS of
charged leptons from glueballs at strong coupling, where $N$ is the
number of color degrees of freedom of the gauge theory. We have done
it by considering the gauge/string duality. We have considered the
AdS$_5 \times S^5$ background with a hard cutoff $z_0=\Lambda^{-1}$,
where $\Lambda$ is the IR confinement scale in the gauge theory. In
the bulk description the initial hadron is represented by a dilaton
with a conformal dimension $\Delta$, while a massless $A_m$ vector
is associated with the perturbation produced by the insertion of the
electromagnetic currents (it can be the $R$-symmetry current) at the
boundary, and it is interpreted as a dual virtual photon.

The DIS high energy limit is when $q^2 \gg \Lambda^2$, where $q$ is
the four-momentum of the virtual photon. On the other hand, for the
AdS/CFT correspondence the gauge theory processes are studied in the
planar limit, and from that it is possible to investigate
corrections in the $1/N$ expansion of the gauge theory. From the
string theory point of view this corresponds to the genus expansion.
In the low energy limit of string theory it becomes the supergravity
loop Feynman diagram expansion.

The idea of this work is to study the compatibility between these
two limits. Our results show that they do not commute. By
considering first the $N\rightarrow \infty$ limit, it leads to the
case where DIS is described by a bulk process with only one
intermediate state which results in structure functions proportional
to $(\Lambda^2/q^2)^{\Delta-1}$. On the other hand, by taking first
the high energy limit $q^2 \gg \Lambda^2$ particle creation is
allowed, and the resulting two-intermediate particle process renders
structure functions proportional to $1/N^2$ and $(\Lambda^2/q^2)$.
In a way this is expected since the high energy limit allows
particle creation.

From first principles we have described the bulk processes that
contribute to the $1/N^2$ corrections to DIS in terms of the
holographic forward Compton scattering (related to DIS via the
optical theorem) with two-particle intermediate states, {\it i.e.}
by calculating the corresponding one-loop Witten diagrams. For this
purpose, we have described the relevant supergravity fluctuations in
terms of an expansion in spherical harmonics on $S^5$, focusing on
dilatons and gravitons, more specifically scalar and vector
fluctuations of the metric, together with their interactions. By
using the interaction terms we have studied the corresponding Witten
diagrams. We have concluded that at order $1/N^2$ and in the DIS
regime of the gauge theory there is only one leading diagram: the
$t$-channel. This specific channel must be considered on both sides
of the cut, together with the sum over all possible intermediate
states. It is the dominant contribution. The incoming hadron splits
into two other hadrons in the IR region, producing a dilaton and a
scalar $s$ with the lowest conformal dimension $\Delta'=2$, each one
carrying a fraction of the incoming hadron momentum. Then, only the
second particle tunnels to the UV region and interacts with the
$A_m$ field. The appearance of this $s$ particle is the reason why
the $t$-channel is the dominant diagram. It leads to further
consequences. In the $N\rightarrow \infty$ limit, the photon strikes
the entire hadron, which implies that $F_1=0$. Beyond this limit,
{\it i.e.} by including the first $1/N^2$ correction, the hadron is
fragmented and the photon interacts only with one of the resulting
particles, which leads to $F_1 \neq 0$. In fact, $F_1$ and $F_2$ can
be explicitly separated in two parts: the first terms of each
structure function are related by the Callan-Gross relation $F^*_2 =
2xF^*_1$, while the second ones give a non-zero contribution to the
longitudinal structure function $F_L \equiv F_2 - 2x F_1$. This
unveils a richer structure for the currents, since both $F_1$ and
$F_2$ are non-vanishing in this limit, which means that the currents
can, in principle, contain spin-$1$, spin-$1/2$ and spin-$0$ fields
inherited from the ${\cal{N}} = 4$ SYM supermultiplet. The expansion
of equation (\ref{FL}) allows one to understand more about the
current structure inside the glueballs at strong coupling. This in
fact holds for any holographic dual pair of theories whose
asymptotic geometry is AdS$_5 \times S^5$.

Also, from the calculation of the amplitude we have obtained the
$q^2$ dependence of $F_L(x, q^2)$ and, within some approximations,
its exact functional form at order $1/N^2$ (\ref{FL2}). It turns out
to be completely consistent with the field theory OPE prediction
discussed by Polchinski and Strassler. Furthermore, we found the $x$
dependence of $F_L \propto x^3(1-x)^2$ which compares well with
phenomenology and lattice-QCD results \cite{Koile:2015qsa}. In
consequence, this represents an explicit example where $q
\rightarrow \infty$ and $N\rightarrow \infty$ limits do not commute.
In addition, the $x$-dependence implies that $F_L$ goes to zero at
$x=0$ and $x=1$ and it is bell-shaped with a maximum at $x \approx
0.6$, as expected. It is also consistent with the fact that, for
some particles (for example the $\pi$-meson) comparison with
experimental results have shown that valence structure functions
behave like $(1-x)^2$ when $x\rightarrow 1$ \cite{Koile:2015qsa}
(and references therein). Note that in previous work we have seen
that the concepts of \textit{valence structure functions} and the
contribution of the \textit{sea of quarks} are related in the
context of holographic calculations with the contributions coming
from the supergravity regime (at $\lambda^{-1/2}\ll x<1$) and those
coming from string theoretical considerations
($\exp(-\lambda^{1/2})\ll x \ll \lambda^{-1/2}$), respectively. We
have also found that $F_1$ turns out to be subleading in the $1/q^2$
expansion. This means that obtaining its explicit form from the
$t$-channel diagram would have been meaningless since contributions
coming from other diagrams could be of the same order.

In addition, we have discussed DIS considering multi-hadron final
states, analyzing the general structure of contributions of higher
order loop expansion under a few assumptions based on the $1/N^2$
case. We have found that the fundamental first steps of the our
previous analysis remain unchanged\footnote{This refers to the steps
we followed in the case of two-particle intermediate states up to
the end of Section 3.1.}. Aside from the possible IR process, where
the hadron splits into multiple particles leading to multi-particle
intermediate states in FCS, the appearance of an $s$ scalar with
conformal dimension $\Delta'=2$ is needed in order to have the
lowest possible $\Lambda^2/q^2$ suppression. This is the particle
that interacts with $A_m$ in the small-$z$ region, leading to an
identical tensor structure. The overall $q$-dependence seems to be
the same in all the $n$-loop cases with $n \geq 1$, implying that
the results of this paper together with the ones in
\cite{Polchinski:2002jw} are the only ones relevant for glueball DIS
at strong coupling, at least in the regime where supergravity
provides an accurate description.

In conclusion, if hadron production is forbidden (the large $N$
limit), the most relevant term in equation (\ref{FL}) is
$f_L^{(0)}$. On the other hand, if hadron production is possible,
$f_L^{(1)}$ becomes the leading one, since the rest of terms have a
structure as shown in figure 6 where the multi-loop with $(n-1)$
hadrons occur in the IR, while a single hadron tunnels towards the
UV of the gauge theory as commented before. Then, the net effect is
similar as having one-loop corrections.

In addition, notice that in the expression for the moments the third
term with the factor $1/N^2$ dominates the expression for
$M_n^{(s)}(q^2)$ when $q^2 \geq \Lambda^2 N^2/(\tau_{\cal
{Q}}-\tau_c)$ for which $\tau_{\cal {Q}}=\tau_c+1$.

Possible extensions can be studied with the techniques presented in
this work. For instance one can consider a different background of
the type AdS$_5 \times {\cal {C}}^5$ (for a compact Einstein
manifold ${\cal {C}}^5$). In this case if the five-dimensional
reduction from type IIB supergravity is known, in principle, one can
calculate the $1/N^2$ corrections in a similar way as described in
this work. In general, one would expect that if there appear
$\alpha'$ and/or $1/N$ corrections to the background, these
corrections may affect the region where the cutoff of the AdS space
is located. In that case, since the loop corrections we study have
the virtual photon interaction with two scalar fluctuation near the
UV, we would expect similar conclusions. Another interesting
possibility from the theoretical point of view is to study DIS
processes in gauge theories in different spacetime dimensions. As an
interesting possibility one could consider the $(0,2)$ theory, and
study the scattering amplitudes by using the AdS/CFT correspondence
in the AdS$_7 \times S^4$ from eleven-dimensional supergravity, and
then to include loop corrections. The consistent dimensional
reduction in that case has been done in \cite{Nastase:1999cb}. Also,
a similar procedure can be thought to be carried out for AdS$_4
\times S^7$ from eleven-dimensional supergravity, for a dual
three-dimensional gauge theory.

~

~

\centerline{\large{\bf Acknowledgments}}

~

We thank Jos\'e Goity, Sergio Iguri and Carlos N\'u\~nez for
valuable comments on the manuscript. The work of D.J., N.K. and M.S.
is supported by the CONICET. This work has been partially supported
by the CONICET-PIP 0595/13 grant and UNLP grant 11/X648.


\newpage

\appendix

\section{Axial gauge calculations}
\subsection{Scalar, vector and tensor solutions}

In this appendix we briefly review the solutions for the different
bosonic fluctuations and their derivations. In the massless cases we
follow the work in \cite{Raju:2011mp}. As we have done in Section 2,
the different fluctuations on the AdS$_5\times S^5$ background can
be expanded in terms of the spherical harmonics of the $S^5$. Once
this is done for each supergravity mode one obtains a series of
massive scalar, vector and tensor fluctuations in AdS$_5$.
Kaluza-Klein masses are given by the eigenvalues with respect to the
angular Laplacian. Moreover, it is useful to work with a complete
set of momentum eigenfunction solutions of the form
$\Phi(x^0,\dots,x^3,z)=e^{ip\cdot x}\Phi^{(p)}(z)$ and focus on the
timelike momentum case.

There are three different Kaluza-Klein towers of scalar
fluctuations, labeled as $s$, $t$ and $\Omega$ in table 1. As we
have seen, they have different Kaluza-Klein mass formulas but in
what follows we will use generically $m \propto R^{-1}$. Like any
scalar massive mode in AdS$_5$ they are defined by the Klein-Gordon
equation
\begin{equation}
(\Box-m^2)\phi(x,z)=0  \Rightarrow \left[z^2\partial^2_z -
3z\partial_z + \left(z^2p^2-R^2m^2\right)\right]\phi^{(p)}(z)=0 \, ,
\end{equation}
with $p^2=\eta_{\mu\nu} p^\mu p^\nu$. In the massless case the
equation is equivalent to
$\partial_m(\sqrt{-g}g^{mn}\partial_n)\phi=0$ and one obtains the
well known solutions
\begin{eqnarray}
\phi^{(p)}(z) &\sim & z^2 J_2(p z) \ \ \textrm{normalizable}, \\
\phi^{(p)}(z) &\sim & z^2 Y_2(p z) \ \ \textrm{non-normalizable},
\nonumber
\end{eqnarray}
where $p\equiv \sqrt{-p^2}$ and $J$ and $Y$ are the Bessel functions
of the first and second kind, respectively. For the diagrams we
study in this work we only need the normalizable modes. Henceforth,
we omit the non-normalizable ones\footnote{The only non-normalizable
solution is the vector perturbation produced by the insertion of the
current at the boundary. This goes exactly as in
\cite{Polchinski:2002jw}.}. When $m^2\neq 0$ the story is similar
and one has
\begin{equation}
\phi^{(p)}(z)\sim z^2 J_{\sqrt{4+R^2m^2}}(pz) \, .
\end{equation}
Note that this implies that the solutions are different for each
type of scalar. In general, for scalar fluctuations the scaling
dimension $\Delta \geq 2$ of the associated operator of the boundary
field theory is given by $m^2=R^{-2}\Delta(\Delta-4)$. Thus, the
Bessel function index is given by $\sqrt{4+R^2m^2} = \Delta-2$. This
is of course the same in any gauge. For completeness we write the
scalar propagator corresponding to the solutions above
\begin{eqnarray}
G (x,z;x',z') &=& \frac{1}{Vol(S^5)R^3} \int \frac{d^4p}{(2\pi)^4}G^{(p)}(z,z')e^{ip\cdot(x-x')} \nonumber \\
&=& -\frac{i}{\pi^3 R^8}\int \frac{d^4p}{(2\pi)^4}\frac{dM^2}{2}
\frac{z^{2} \, J_{\Delta-2}(M z) z'^{2} J_{\Delta-2}(M
z')e^{ip\cdot(x-x')}}{p^2 + M^2- i \epsilon} \, . \label{propagator}
\end{eqnarray}
This propagator is easily obtained by solving the
equation\footnote{This is in five dimensions, but we also have to
integrate on the sphere.}
\begin{equation}
\Box G (x,z;x',z') = \frac{i}{\sqrt{-g_{AdS_5}}}\delta^4(x-x')
\delta(z-z') \, ,
\end{equation}
in Fourier space (for the first four coordinates) \cite{Raju:2011mp}
using the identity
\begin{equation}
\int dz \, z \, J_\nu (M z) J_\nu(M' z) = \frac{1}{M} \, \delta
(M-M') \, .
\end{equation}
Let us recall that there is a cutoff $z_0$ in the AdS space where
the solutions have to vanish. This means that $p$ has to be such
that the product $p z_0$ is a zero of the Bessel function. This
holds for all the normalizable fields we consider.

For vector fields in AdS$_5$ (that we generically denote $A_{m}$)
one has to solve the Einstein-Maxwell equation after fixing some
gauge degrees of freedom. The axial gauge is defined by imposing
$A_z=0$. Thus, after separating variables one finds
\begin{equation}
\left[z^2\partial_z^2 - z\partial_z +
(z^2p^2-R^2m^2)\right]A^{(p)}_\mu=0 \ \ \textrm{and} \ \
\eta^{\mu\nu}p_\mu A_\nu^{(p)} = 0 \, ,
\end{equation}
where the last equation only holds for normalizable modes. This
system has the following massive solutions
\begin{equation}
A_\mu^{(p)}(z)\sim \epsilon_\mu \,z\, J_{\sqrt{1+R^2m^2}}(pz) \
\textrm{with} \ p\cdot \epsilon=0 \, .
\end{equation}
The definition $m^2 = R^{-2}(\Delta-1)(\Delta-3)$ for vector
fluctuations in the context of the AdS/CFT correspondence leads to
an index of the form $\sqrt{1+R^2m^2}=\Delta-2$. The only difference
with the scalar case is the power of the $z$ factor.

Now, let us consider the tensor fluctuations $h_{\mu\nu}$. There is
only one Kaluza-Klein tower of these states, and among them the
massless one corresponds the AdS$_5$ graviton. In this case, the
axial gauge is defined by $h_{\mu z}=0$, which leads to important
simplifications and, as in the vector case, it selects the
transversally polarized solutions for
$h_{\mu\nu}$\footnote{Formally, there is also a mode associated with
$h_{zz}$, however this is not an independent degree of freedom since
the trace $h^{\mu}_{\mu}$ is already included in the scalar
fluctuations.}. After some algebra, one finds that the equations of
motion are given by
\begin{equation}
\left[z^2\partial_z^2 + z\partial_z +
\left(z^2p^2-4-R^2m^2\right)\right]h_{\mu\nu}^{(p)}=0\ \
\textrm{and} \ \ \eta^{\mu\nu}p_{\mu}h_{\mu\sigma}^{(p)} = 0 \, ,
\end{equation}
where we are only left with symmetric traceless perturbations. Thus,
the solutions are of the form
\begin{eqnarray}
h_{\mu \nu}^{(p)}(z)\sim E_{\mu \nu}\, J_{\sqrt{4+R^2m^2}}(pz) \
\textrm{with} \
\eta^{\nu\sigma}p_{\nu}E_{\sigma\mu}=\eta^{\nu\sigma}p_{\nu}E_{\mu\sigma}=0
\ ,
\ \eta^{\mu\nu}E_{\mu\nu}=0 \ , \ E_{\mu \nu} = E_{\nu\mu}, \nonumber \\
\end{eqnarray}
and since the $m^2(\Delta)$ equation for tensor modes is the same as
in the scalar case, while the index is $\Delta-2$ with a different
$z$ factor.

The canonical normalization condition for scalars involves the
cutoff $z_0$ and is given in \cite{Polchinski:2001tt, Gao:2014nwa},
where it is shown that for a field of the form $\phi = e^{ip\cdot
x}f(z)Y(\Omega_5)$, canonical quantization implies the normalization
condition
\begin{equation}
\int_0^{z_0} \, dz \, d\Omega_5 \, \omega(z) \, \sqrt{g_{zz}g_{S^5}}
\,  |f(z) \, Y(\Omega_5)| = 1 \, ,
\end{equation}
where $\omega(z)=(R/z)^{2}$ is the warp factor multiplying
$\eta_{\mu\nu}dx^{\mu}dx^{\nu}$ in the metric, and in a more general
context $g_{zz}g_{S^5}$ should be replaced by the determinant of the
part of the metric corresponding to the rest of the coordinates.
Assuming that the angular part of the solution is normalized as
\begin{equation}
\int d\Omega_5  \, \sqrt{g_{S^5}} \, |Y(\Omega)|^2=1 \, ,
\end{equation}
and using the fact that our solutions vanish at $z=z_0$, which means
that the $J_{\Delta-2}(p z_0)=0$, the normalization constant is
\begin{equation}
c = \frac{\sqrt{2}}{z_0 R^4 |J_{\Delta-1}(p z_0)|}.
\end{equation}
By taking into account that $|A^{\mu}A_\mu|\sim z^2|A_{\mu}|^2$ and
$|h^{\mu\nu}h_{\mu\nu}|\sim z^4|h_{\mu\nu}|^2$ the vector and tensor
normalizations are obtained in a similar way.

\subsection{Details of the planar limit}

As we have seen, in the axial gauge we set $A_z=0$, and after
proposing a solution of the form $A_{\mu} =c_{\mu} e^{i k\cdot x}
f(z)$ the Einstein-Maxwell equations of motion for the massless
vector coming from the boundary become
\begin{equation}
i q \cdot \partial_z A=0 \ , \
\partial^2_z A_\mu-\frac{1}{z}\partial_z A_\mu - q^2 A_\mu + q_{\mu} q \cdot A=0 \, ,
\end{equation}
where the contraction stands for $v \cdot w = \eta^{\mu\nu} v_\mu
w_\nu$. The first equation implies that $q\cdot A$ is a constant in
terms of the $z$ variable. For normalizable modes, as
$A_\mu(z\rightarrow 0)\rightarrow 0$ this simply implies that $q
\cdot A =0$, and we can forget about it in the second equation, as
we have done before \footnote{This is important since this constant
term would yield much more complicated solutions in the massive
case.}. However, if we want $A_{\mu}$ to describe an $R$-current
excitation coming from the boundary, we can no longer ignore this
constant because of the boundary condition
\begin{equation}
A_\mu (z\rightarrow 0) \rightarrow n_{\mu} e^{iq\cdot x} \Rightarrow
q\cdot A|_{z=0} = q \cdot n \, e^{iq\cdot x} = \textrm{const.} \, .
\end{equation}
The full non-normalizable solution takes the form
\begin{equation}
A_{\mu} = \left[c_{\mu} q z K_1(qz) + \frac{(q\cdot
A)q_{\mu}}{q^2}\right] e^{iq\cdot x} \, ,
\end{equation}
and imposing the boundary condition leads to
\begin{equation}
c_\mu = n_{\mu} - \frac{(q\cdot n)q_{\mu}}{q^2} \, .
\end{equation}
Recall that in the Lorentz gauge one obtains $c_{\mu}=n_{\mu}$ (and
$A_z \neq 0$). Now, writing the current as $J_{m} = i {\cal {Q}}
(\phi_I
\partial_m \phi_X^{\star} - \phi_X^{\star}\partial_m \phi_I)$ the
interaction action evaluated on-shell in the gauge that we consider
is
\begin{eqnarray}
S_{A\phi\phi} &= & i {\cal {Q}} \int d^{10}x \, \sqrt{-g} A^m J_m
= i {\cal {Q}} \int d^{10}x \, \sqrt{-g} A^\mu J_\mu \nonumber\\
&=& i {\cal {Q}} \int d^{10}x \, \sqrt{-g} \phi_I \phi_X^{\star}
A^\mu \left(2P_{\mu} + q_{\mu}\right) \, , \nonumber
\end{eqnarray}
which represents a term coming from the Bessel function and another
from the $z$-constant terms of $A_\mu$.

The former gives exactly the $z$ integrand of the Lorentz case
$z^{\Delta} J_{\Delta-2}(s^{1/2}z)K_1(qz)$, and noting that the
contraction is
\begin{equation}
c^{\mu} \left(2P_{\mu} + q_{\mu}\right) = \left(n_{\mu} -
\frac{(q\cdot n)}{q^2} q_{\mu}\right)\left(2P^{\mu} + q^{\mu}\right)
= 2 n \cdot \left(P + \frac{q}{2x} \right),
\end{equation}
it leads to the same contribution as in \cite{Polchinski:2002jw}.
This means that the other term must vanish, and it is what happens.
Since $A$ does not fall down rapidly with $z$ in the bulk, one
cannot use the asymptotic behavior for the ingoing state, which
means that the $z$-integral is of the form
\begin{eqnarray}
&&\int_0^{z_0} dz\, z J_{\Delta-2}(s^{1/2}z)J_{\Delta-2}(P z) =
\nonumber
\\ &&\frac{z_0}{s-P^2} \left[s J_{\Delta-3}(s^{1/2}
z_0)J_{\Delta-2}(P z_0)-P J_{\Delta-2}(s^{1/2} z_0)J_{\Delta-3}(P
z_0)\right] \, .
\end{eqnarray}
$J_{\Delta-2}$ must vanish at $z_0$, which proves that the constant
term that appears in the axial gauge does not contribute to the
structure functions.

\section{Double and triple Bessel function integrals}

The following are known definite integrals that we use in this work.
They come from the Bessel functions $J_\mu (x)$ and $K_\mu (x)$
which are present in all the non-asymptotic solutions of free fields
in the AdS$_5\times S^5$ space.
\begin{itemize}
\item For vertices between three normalizable states we can use a
semi-empirical distribution \cite{Auluck:2012}:
\begin{eqnarray}
\int_0^{\infty} z J_{\Delta_1-2}(M_1 z)J_{\Delta_2-2}(M_2 z)
J_{\Delta_3-2}(M_3 z) dz \approx \ \ \ \ \ \ \ \ \ \ \ \ \   \ \ \ \
\ \ \ \ \ \ \ \ \ \ \ \  \ \ \ \ \ \ \ \ \ \   \nonumber
\\
 \left[ \cos \left(\frac{\pi (\Delta_3 -\Delta_1 -\Delta_2+2}{2}  \right)
 + \sin \left(\frac{\pi (\Delta_3 -\Delta_1 -\Delta_2+2   )}{2} \right) \right] \nonumber
 \ \ \ \ \  \ \\ 
 \times  \frac{\delta (M_3 -(M_1+M_2))}{2\sqrt{M_1 M_2}} \ \ \ \ \ \ \ \ \ \ \
 \ \ \ \  \  \ \ \ \ \ \ \  \ \ \ \ \ \ \  \ \ \ \ \  \ \ \ \ \ \ \  \ \  \ \  \ \ \ \ \ \ \ \ \ \ \nonumber
\\
+ \left[ \cos \left(\frac{\pi (\Delta_3 -|\Delta_1 -\Delta_2| +2   )}{2} \right)
+ \sin \left(\frac{\pi (\Delta_3 -|\Delta_1 -\Delta_2|+2   )}{2} \right) \right]
\nonumber \\
\times \frac{\delta (M_3 -(|M_1-M_2|)}{2\sqrt{M_1 M_2}}  \, . \ \ \
\ \ \ \ \ \ \ \ \ \ \ \ \ \ \  \ \ \ \ \ \ \  \ \ \ \ \ \ \ \  \ \ \
\ \ \ \  \ \ \ \ \ \ \  \ \ \ \ \ \ \
\label{semiemp}
\end{eqnarray}
\item For the same vertices with two equal states
\begin{eqnarray}
\int_0^\infty z J_0(a z)J_{\nu}(b z)J_{\nu}(c z) dz =
\frac{1}{bc\sqrt{2\pi\sin(v)}}P_{\nu-\frac{1}{2}}^{\frac{1}{2}}\left(\cos(v)\right)
\, ,
\end{eqnarray}
if $|b-c|<a<b+c$ or zero otherwise. $P_\alpha^\beta(x)$ represents
the associated Legendre function and we have defined $\cos(v)\equiv
\frac{b^2+c^2-a^2}{2bc}$.

\item For vertices between two normalizable states and a non-normalizable
perturbation coming from the boundary
\begin{eqnarray}
\int_0^{\infty}z^{\rho-1} J_{\lambda}(az)J_{\mu}(bz)K_{\nu}(cz)dz =
\frac{2^{\rho-2}a^{\lambda}b^{\mu}c^{-\rho-\lambda-\mu}}{\Gamma(\lambda+1)
\Gamma(\mu+1)}\Gamma\left(\frac{\rho+\lambda+\mu-\nu}{2}\right)\nonumber \ \ \ \ \ \ \ \ \ \ \ \ \ \  \\
\times \Gamma\left(\frac{\rho+\lambda+\mu+\nu}{2}\right)F_4
\left(\frac{\rho+\lambda+\mu-\nu}{2},\,\frac{\rho+\lambda+\mu+\nu}{2};
\lambda+1 ,\, \mu+1;-\frac{a^2}{c^2},-\frac{b^2}{c^2}\right)
\label{intJJK} \nonumber \\
&&
\end{eqnarray}
where in this equation $F_4$ is the fourth Appell series of
hypergeometric functions. This formula is valid if $\textrm{Re}(
\rho+\mu+\lambda)>\textrm{Re}(\nu)$ and $\textrm{Re}(c)>
|\textrm{Im}(a)|+|\textrm{Im}(b)|$.
\item For the same vertices with a normalizable state which is
approximated by its asymptotic expansion
\begin{eqnarray}
\int_0^{\infty}z^{\rho}K_{\mu}(az)J_{\nu}(bz)dz =
2^{\rho-1}\left(\frac{b}{a}\right)^{\nu}a^{-\rho-1}
\frac{\Gamma\left(\frac{\nu +\rho+\mu+1}{2}\right)
\Gamma\left(\frac{\nu +\rho-\mu+1}{2}\right)}{\Gamma(\nu+1)}\nonumber \\
\times F\left(\frac{\nu +\rho+\mu+1}{2},\,\frac{\nu +\rho-\mu+1}{2},
\ \nu+1 ,\ -\frac{b^2}{a^2}\right) \, . \label{intJK}
\end{eqnarray}
This equation is valid if $\textrm{Re}(a\pm ib)>0$ and
$\textrm{Re}(\nu+\lambda+1)> |\textrm{Re}(\mu)|$.

For $\rho=\Delta$, $\nu=\Delta-2$, $\mu=1$, $a=q$ and $b=\sqrt{s}$,
we have that
$F(\Delta,\Delta-1,\Delta-1,-\frac{s}{q^2})=(1+\frac{s}{q^2})^{-\Delta}$
and we recover the result of equation (\ref{JJK}).

\end{itemize}

\newpage


\end{document}